\documentclass{aa}

\usepackage[varg]{txfonts}
\usepackage{graphicx}
\usepackage{xcolor}
\usepackage{mathrsfs}
\usepackage{booktabs}
\usepackage{tabularx}
\usepackage{multicol}
\usepackage{subcaption}
\usepackage{caption}

\RequirePackage{nameref}

\begin{document}

\title{Cosmic magnification on multi‑catalogue Herschel\thanks{{\it Herschel} is an ESA space observatory with science instruments provided by European-led Principal Investigator consortia and with important participation from NASA.} submillimetre galaxies.}
\titlerunning{MagBias Herschel}
\authorrunning{Fern{\'a}ndez-Fern{\'a}ndez R. et al.}

\author{
  Fern{\'a}ndez-Fern{\'a}ndez R.\inst{1,2} \and
  Cueli M. M.\inst{3,4} \and
  Gonz{\'a}lez-Nuevo J.\inst{1,2} \and
  Bonavera L.\inst{1,2} \and
  Crespo D.\inst{1,2} \and
  Goitia E.\inst{1} \and
  Casas J. M.\inst{1,2,5,6} \and
  Cano J. A.\inst{1,2} \and
  Migliaccio M.\inst{7,8}
}

\institute{
  Departamento de F\'{i}sica, Universidad de Oviedo, C. Federico Garc\'{i}a Lorca 18, 33007 Oviedo, Spain \and
  Instituto Universitario de Ciencias y Tecnolog\'{i}as Espaciales de Asturias (ICTEA), C. Independencia 13, 33004 Oviedo, Spain \and
  SISSA, Via Bonomea 265, 34136 Trieste, Italy \and
  IFPU - Institute for Fundamental Physics of the Universe, Via Beirut 2, 34014 Trieste, Italy \and
  Instituto de Astrof\'{i}sica de Canarias, E-38200 La Laguna, Tenerife, Spain \and
  Universidad de La Laguna, Departamento de Astrof\'{i}sica, E-38206 La Laguna, Tenerife, Spain \and
  Dipartimento di Fisica, Università di Roma Tor Vergata, Via della Ricerca Scientifica 1, 00133 Roma, Italy \and
  INFN Sezione di Roma2, Università di Roma Tor Vergata, Via della Ricerca Scientifica 1, 00133 Roma, Italy
}

\date{}
\abstract
{Submillimetre galaxies (SMGs) are exceptional background sources for magnification‑bias studies, but the inherently limited sky coverage in the submillimetre (sub-mm) band constrains their statistical power. Beyond H‑ATLAS, Herschel has produced additional sub-mm catalogues, although not optimised for spatial statistical lensing analyses. 
}
{Our goal is to improve the current cosmological constraints derived from SMG magnification bias by exploiting the full sub-mm sky surveyed by Herschel.
}
{ We enlarged the SMG sample by incorporating other Herschel catalogues that overlap with SDSS spectroscopic lens samples. Random catalogues were generated via kernel density estimation to compute the cross‑correlation functions. Markov Chain Monte Carlo techniques were then employed to extract both astrophysical and cosmological parameters for each individual catalogue and for the combined dataset.
}
{This work delivers the first detection of magnification bias in SMGs outside the H-ATLAS catalogue, further supporting the robustness of the observable. Individual \textit{Herschel} catalogues provide central values for $\Omega_m$ and $\sigma_8$ that are reasonable, but their constraints are affected by large uncertainties. In contrast, the combined analysis—dominated by the statistically more powerful H-ATLAS dataset—yields constraints consistent with the $\Lambda$CDM model. In particular, it significantly improves the determination of the matter density parameter, finding $\Omega_m = 0.30^{+0.05}_{-0.07}$, $\sigma_8 = 0.80 \pm 0.07$, and $h < 0.80$. These mean values are better aligned with the \textit{Planck} 2018 results than those obtained in previous non-tomographic magnification bias studies. 
}
{SMGs prove to be excellent background sources for magnification bias, but the limited sky coverage in the sub-mm remains a major constraint. Wider surveys specifically designed for lensing would make it possible to probe larger angular scales via cross‑correlation, ultimately delivering more competitive constraints. 
}

\keywords{galaxies: high-redshift -- submillimeter: galaxies -- gravitational lensing: weak -- cosmology: cosmological parameters -- methods: data analysis}

\maketitle

\section{Introduction}

Gravitational lensing, the deviation of light-ray pathways in the presence of large distributions of matter (baryonic and dark), has been thoroughly exploited to study cosmology over the last decades. Gravitational lensing induces several measurable effects on distant sources, due to the apparent magnification and distortion produced by the deflection of light \citep{SCH92}.

An interesting gravitational lensing effect in the weak lensing regime is magnification bias. It produces an excess (or defect) of background (high redshift) source number counts in the vicinity of foreground (low redshift) mass structures, such as galaxies or clusters of galaxies, when a flux threshold is applied (see \cite{Bar01} for a detailed explanation). This cross-correlation function (CCF) arises from the deflection of light rays from high-redshift sources as they traverse the low-redshift mass density field. It is dependent on the mass distribution, as well as on cosmological distances and galaxy halo properties. This makes magnification bias a probe for Cosmology complementary but independent of lensing collaborations which focus on shear, such as Euclid \citep{EUCLID25}, the European Southern Observatory Kilo-Degree Survey\footnote{http://kids.strw.leidenuniv.nl} (KiDS), the Dark Energy Survey\footnote{https://www.darkenergysurvey.org} (DES), and the \textit{Planck} Collaboration\footnote{https://www.cosmos.esa.int/web/planck/pla}. 

Magnification bias has been successfully measured in previous works among different background and foreground sources: low-redshift galaxies and high-redshift quasars \citep[e.g.,][]{SCR05, MEN10}, foreground \textit{Herschel} sources and Lyman-break galaxies \citep[e.g.,][]{HIL13}, cosmic microwave background and other source distributions \citep[e.g.,][]{BIA15, BIA16}, among others. In particular, submillimetre galaxies (SMGs) have been exploited due to their outstanding properties as background sources for lensing, namely their steep luminosity function, their high redshift and faint optical emission \citep{BLA96,  Neg07, NEG10, FU12,  BUS12, GON12, BUS13,  WAR13, CAL14, NAY16, GON17, NEG17, Bak20}. Furthermore, magnification bias-induced CCFs between SMGs and different foreground samples have already been studied as cosmological probes in the past decade, in single redshift bin analyses \citep[][]{GON17, BON20, CUE21, GON21, BON23, CUE24, FER24} and also using a redshift tomographic approach, which shows promising results as a way to improve the constraints on the parameters under study \citep[][]{BON21, CUE22, BON24}.

However, a significant handicap of SMG lensing studies is the limited sky coverage, which compromises the precision of the results. All the aforementioned studies have used the SMGs detected in the Herschel Astrophysical Terahertz Large Area Survey \citep[H-ATLAS; ][]{EAL10, SMI17}, that encompasses five different sky fields. Three of them overlap with the Galaxy and Mass Assembly \citep[GAMA;][]{DRI11} survey, located along the celestial equator at right ascensions of 9, 12, and 14.5 hours (G09, G12, and G15). The remaining fields are located at the North and South Galactic Poles (NGP and SGP).  

The aim of this study is to enlarge the high-redshift SMG sample using additional Herschel regions in order to improve the precision of astrophysical and cosmological constraints derived from previous H-ATLAS-based magnification bias studies. However, there is a limited amount of observational data available in the submillimetre sky. ESA's Herschel Space Observatory functioned primarily as a space observatory rather than a dedicated survey mission, and only a few regions cover areas large enough to be considered surveys with the minimum areas required for CCF measurements (typically at least 4 deg$^2$). Other than H-ATLAS, the collaborations that produced the largest-area regions, and which will be used in this work to prepare the background samples, are: the Herschel Multi-tiered Extragalactic Survey \citep[HerMES; ][]{HerMES}  the Herschel Virgo Cluster Survey \citep[HeVICS; ][]{HeVICS}, and the Herschel Stripe 82 Survey \citep[HerS; ][]{HerS}. Additionally, the Herschel Spectral and Photometric Imaging Receiver (SPIRE) Point Source Catalogue \citep[HSPSC\footnote{European Space Agency, 2017, Herschel SPIRE Point Source Catalogue, Version 1.0. https://doi.org/10.5270/esa-6gfkpzh}; ][]{PIL10}, which was produced \textit{a posteriori} from Herschel observations with the aim of providing a homogeneous source detection, was also employed as a background catalogue.

Regarding the lenses, the GAMA catalogue was used for the H-ATLAS studies, but it does not cover the new sky regions. For this reason, overlapping regions of galaxies with spectroscopic redshift corresponding to the fifth phase of the Sloan Digital Sky Survey \citep[SDSS-V; ][]{SDSSV}  were used as the foreground sample. To ensure no redshift overlap with the background, a redshift cut was applied between 0.05 and 0.8. Individual CCF and Markov Chain Monte Carlo (MCMC) analyses were performed for each background catalogue, followed by a joint analysis across the non-overlapping sky fields. This approach aimed to maximise the precision achievable with existing data and to evaluate the constraining power of magnification bias at present.

This article is divided into the following sections. The different datasets used as sources and lenses are presented in Section \ref{sec:method}, as well as the methodology employed to obtain the CCF and parameter estimates. Section \ref{sec:results} summarises the results of the measurements. First, a proof of concept of the new methodology is shown \citep[by applying the new techniques to the same dataset used in][]{CUE24}, followed by the CCF and MCMC results for the new sky zones. Finally, the main conclusions of the study are summarised in Section \ref{sec:concl}. The appendices contain additional material. A brief overview of the theoretical framework on weak lensing-induced magnification bias can be found in \ref{app:framework}, while the sky coverage and examples of data quality (in terms of visual homogeneity) are provided in \ref{app:skyzones}. Appendices \ref{app:corner_plots} and \ref{app:data_tables} provide the complete marginalised posterior distributions and parameter constraints from the different MCMC analyses, complementing the main text where only the key results are presented.


\section{Methodology}
\label{sec:method}
\subsection{Data}
\label{subsec:data}

A key aspect of previous studies on magnification bias involved measuring the CCF between galaxies from the GAMA survey and SMGs detected by H-ATLAS. For more comprehensive details on the catalogues and methodologies see \citet{CUE24}. Their results provide a reference point and are incorporated into the joint analysis presented in this paper.

The primary goal of this work is to expand the sample of SMGs analysed in previous works by incorporating all available catalogues overlapping with the SDSS, which serves as the foreground sample (lenses). However, apart from the H-ATLAS survey, there are very few sky zones that have been surveyed in the sub-mm sky, particularly  above 4 deg$^2$, which is the minimum area used in this study. Candidate catalogues with SDSS overlap include HSPSC, HerMES, HeVICS, and HerS. However, while H-ATLAS was designed for statistical analysis and, as such, considerable attention was dedicated to ensuring the homogeneity and completeness of the galaxy catalogues, these other Herschel catalogues were not prepared with that purpose in mind, resulting in inherent inhomogeneities due to variations in catalogue construction, scanning strategies, and data processing techniques. To address these discrepancies, as a first step sky regions were meticulously selected through visual inspection to minimise inhomogeneities, as no systematics masks were available. Residual effects related to catalogue construction were managed by generating randoms that simulate the scanning strategies through kernel density estimation, as outlined in the subsequent section. 

A full-sky picture showing the Herschel sky coverage can be found in Fig. \ref{fig:sky_zones}, and examples of the aforementioned inhomogeneities presented in the catalogues are shown in Fig. \ref{fig:zone-examples}.

Photometric redshifts were estimated for all background sources in the final catalogues with valid detections in all three SPIRE bands. The method, detailed in \citet{GON17} and \citet{BON19}, involves a $\chi^2$ minimization fitting procedure that compares the observed SPIRE photometry with the spectral energy distribution (SED) of SMM J2135-0102, commonly referred to as the "Cosmic Eyelash", a gravitationally lensed starburst galaxy at $z=2.3$. This SED template was selected based on its strong empirical agreement with Herschel-selected high-redshift sources \citep{LAP11, GON12, IVI16}. For sources with incomplete photometry, redshift estimates were omitted. The resulting redshift distribution is shown in Fig. \ref{fig:z_dist}.

\subsubsection{HSPSC (base)}
\label{subsec:HSPSC_base}
To construct one of the background samples employed in our cross-correlation analysis, we utilized the Herschel/SPIRE Point Source Catalogue (HSPSC), a homogeneous photometric dataset compiled from 6878 individual SPIRE observations. The SPIRE instrument onboard the Herschel Space Observatory mapped approximately 9\% of the sky in three broad submillimeter bands centered at 250, 350, and 500 $\mu$m, with beam full-width at half maximum (FWHM) values of 17.9", 24.2", and 35.4". The catalogue includes data from both scientific and calibration programs, extracted through a uniform and robust pipeline optimized for source reliability. The final release excludes regions strongly affected by Galactic foregrounds and provides separate source tables for each band, focusing primarily on point sources, though slightly extended objects are also included. 

To obtain a unified photometric catalogue suitable for redshift estimation and statistical analysis, we performed a cross-matching procedure between the three individual HSPSC band tables. The process began with the 500 $\mu$m band, which has the largest positional uncertainty, and served as the base catalogue. We cross-matched this set with the 350 $\mu$m catalogue using a matching radius equivalent to the 500 $\mu$m beam size. This produced an intermediate catalogue comprising sources detected at both 500 and 350 $\mu$m, as well as those detected only at 500 $\mu$m. Sources detected only at 350 $\mu$m with no counterpart at 500 $\mu$m were excluded (they are considered to be lower-redshift sources, $z<1$). This catalogue was then cross-matched with the 250 $\mu$m catalogue, applying a search radius based on the 350 $\mu$m beam FWHM (when available). To ensure consistent spectral coverage, sources exhibiting detections at 500 and 250 $\mu$m but lacking a counterpart at 350 $\mu$m were also discarded. Sources detected only at 250 $\mu$m or only at 350 $\mu$m were discarded, as they are likely to be spurious sources or lower-redshift ($z<1$) galaxies outside our selection criteria for high-redshift background candidates.

After visual scrutiny to identify homogeneous regions, five different sky patches exclusive to the HSPSC catalogue were selected:"Cosmos", "ELAIS N2", "bootes", "HS82a" and "HS82b". Sources with flux density over 30 mJy and signal-to-noise ratio (SNR) bigger than 4 in any band were selected. In addition, a redshift cut between 1.2 and 4 was introduced to ensure no redshift overlap with the foreground sources. This selection step also implies that all galaxies without photometry in all the SPIRE bands are discarded. This amounts to a total area of 95.4 deg$^2$ and 4100 background sources, with a median redshift of $\langle z \rangle=2.26$. The lenses consist of 29,000 spectroscopically identified SDSS galaxies of medium redshift $\langle z \rangle=0.51$.

\subsubsection{HSPSC (extended)}
\label{subsec:HSPSC_extended}

The sky regions of HSPSC base were expanded by incorporating additional zones from the HSPSC catalogue, specifically HeVICS, HerS, XMM, and Lockman-SWIRE. Although these additional sky zones are also present in the other catalogues (HeVICS, HerS and HerMES), they constitute a sample extracted from a single catalogue with homogeneous detection criteria and provide us with the opportunity to test the robustness of the measurements.  This extension of the HSPSC base results in a total surveyed area of 239 deg$^2$ and includes 12,000 sources, with a mean redshift of  $\langle z \rangle=2.26$. As lenses, 66,700 spectroscopically identified galaxies with a mean redshift of $\langle z \rangle = 0.50$ from SDSS were selected.

\subsubsection{HerMES}
\label{subsec:HerMES}

The Herschel Multi-tiered Extragalactic Survey \citep[HerMES\footnote{Dataset DOI: 10.26131/IRSA78.},][]{HerMES} is a legacy programme designed to map a series of nested fields totalling 380 deg$^2$. For this work, the XMM, Lockman-SWIRE, and ELAIS N2 fields were selected. These fields were observed using Herschel's SPIRE at 250, 350, and 500 microns. After selecting the overlapping area with SDSS and conducting a visual scrutiny and selection of the most homogeneous sky regions, the total area covered is 75 deg$^2$.

Sources were selected based on robust detection criteria, requiring a SNR greater than 4 at either 250 or 350 $\mu$m, along with a reliable redshift measurement ($z/\sigma_z>2$). Additionally, we applied a flux density threshold of 30 mJy and restricted the sample to the redshift range $1.2 < z < 4$. The final sample consists of 19700 galaxies with a median redshift of $\langle z \rangle$=1.79.

The potential lenses were selected from the overlapping area covered by SDSS-V DR18. A total of 15500 galaxies with a median redshift of $\langle z \rangle$ = 0.50 were used as foreground objects.

\subsubsection{HeVICS}
\label{subsec:HeVICS}
The Herschel Virgo Cluster Survey \citep[HeViCS\footnote{Dataset DOI: 10.26131/IRSA70.},][]{HeVICS} is a survey covering approximately 55 deg$^2$ of the nearby Virgo galaxy cluster, conducted with the Herschel Space Observatory using the PACS and SPIRE instruments in parallel mode. It provides wavelength coverage across five bands, ranging from approximately 100 to 600 microns. The final selection consists of a visually homogeneous area of 47.5 deg$^2$. The selection criteria consisted of a redshift between 1.2 and 4, a redshift SNR>2, and a flux density exceeding 30 mJy. This resulted in 1895 sources with a median redshift of $\langle z \rangle = 2$.

The lenses were selected from the overlapping area covered by SDSS. A total of 18,200 galaxies with a median redshift of $\langle z \rangle$ = 0.46 were identified as foreground objects.

\subsubsection{HerS}
\label{subsec:HerS}

The Herschel Stripe 82 Survey \citep[HerS,][]{HerS} is a catalogue derived from observations at 250, 350, and 500 $\mu$m, obtained using the SPIRE instrument aboard the Herschel Space Observatory. HerS covers an area of 79 deg$^2$ along the SDSS Stripe 82, reaching average depths of 13.0, 12.9, and 14.8 mJy beam$^{-1}$ (including confusion) at 250, 350, and 500 $\mu$m, respectively. After visual inspection, the final selected area comprised 38 deg$^2$. Galaxies were chosen based on a redshift range between 1.2 and 4, with reliable redshift detections (SNR > 2) and a strong detection (SNR > 4) at either 250 or 350 $\mu$m. This resulted in a sample of 3700 background galaxies with a median redshift of $\langle z \rangle$ = 1.75.

The lenses were selected from the overlapping region covered by SDSS. A total of 18,200 galaxies, with a median redshift of $\langle z \rangle$ = 0.49, were identified as foreground objects.

\begin{figure}[ht]
\includegraphics[width=0.5\textwidth]{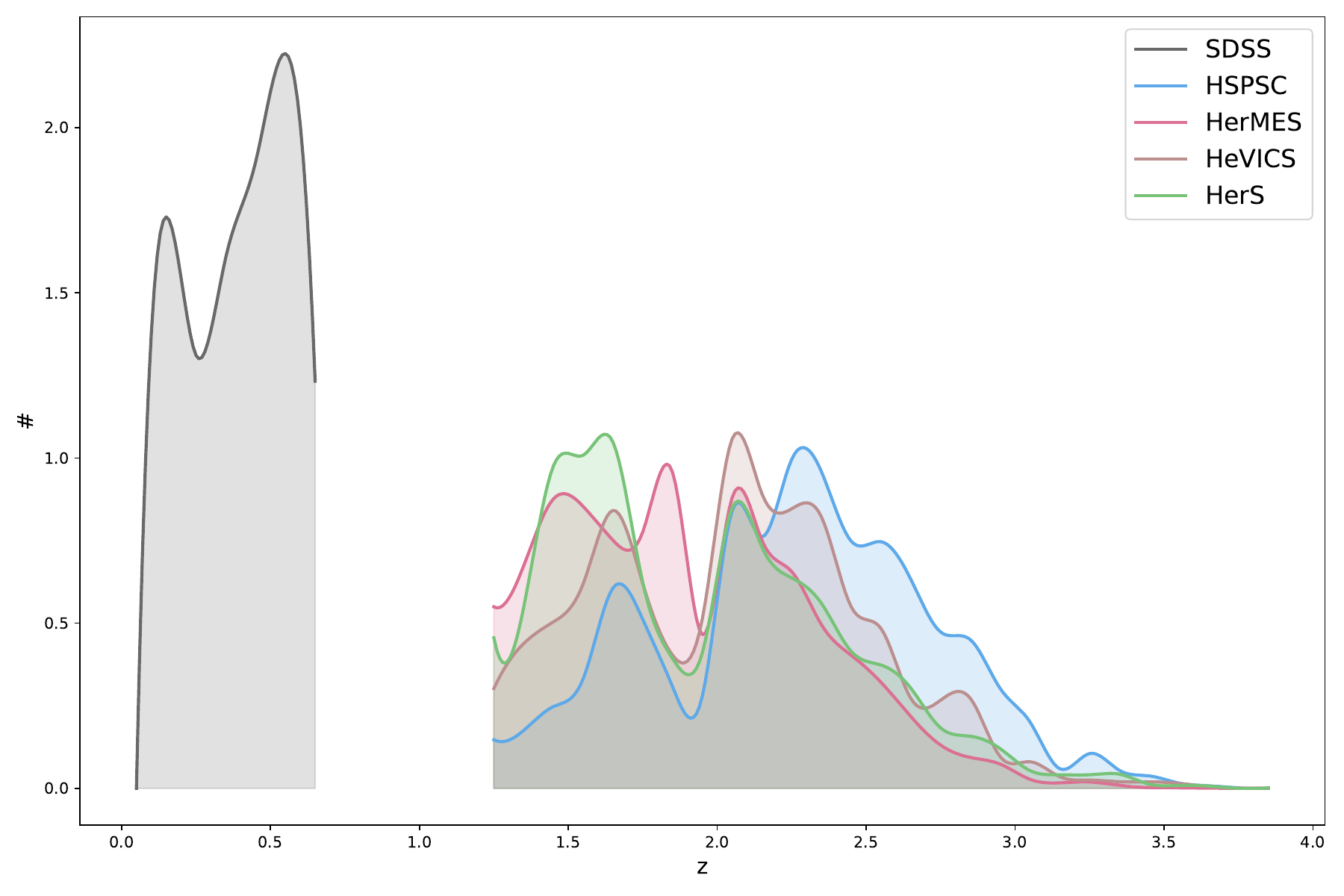}
 \caption{Redshift distributions of the foreground lenses (SDSS, in grey) and the different background SMG samples: HSPSC is depicted in blue, HerMES in pink, HeVICS in brown and HerS in green. The redshift distribution across different fields within a given catalog is very robust.}
 \label{fig:z_dist}
\end{figure}

A summary of the background samples can be found in Table  \ref{tab:catalogues}, and a plot depicting their (smoothed) redshift distributions are shown in Fig.\ref{fig:z_dist}, where the SDSS redshift distribution is shown in grey, and HSPSC, HerMES, HeVICS and HerS are depicted in blue, pink, green and brown, respectively.  

\begin{table}[ht]
  \caption{Summary of the data samples.}
  \label{tab:catalogues}
  \centering
  \small
  \begin{tabularx}{\columnwidth}{cccccc} 
    \hline\hline
    Sample & \multicolumn{2}{c}{HSPSC} & HerMES & HeVICS & HerS\\
           & base & extended & & & \\ 
    \midrule
    Sources & 4100 & 12000 & 19700 & 1895 & 3700 \\
    Lenses & 66700 & 29000 & 15500 & 18200 & 18200 \\
    $\langle z \rangle$ & 2.26 & 2.26 & 1.79 & 2.00 & 1.75 \\
    Area (deg$^2$) & 239 & 95 & 75 & 47.5 & 38 \\
    \hline\hline
  \end{tabularx}
\end{table}


\subsection{Measurements}
\label{subsec:measurements}

\subsubsection{Cross-correlation function estimator}
\label{subsubsec:estimator}
CCF and covariance matrices were measured for each foreground-background catalogue pair following a variation of the method thoroughly discussed in \cite{GON23}. This approach, which has been successfully applied in previous works \citep{BON24, CUE24, FER24} using SMGs from H-ATLAS as the background, aggregates all foreground-background galaxy pairs across all the different sky fields in the catalogue, producing a single measurement. These pairs are then used to compute the CCF with a modified version of the \citet{LAN93} estimator, as adapted by \citet{HER01}: 

\begin{equation}
    \label{eq:w_fb_obs}
    \hat{w}_{\text{fb}}(\theta)=\frac{\rm{D}_f\rm{D}_b(\theta)-\rm{D}_f\rm{R}_b(\theta)-\rm{D}_b\rm{R}_f(\theta)+\rm{R}_f\rm{R}_b(\theta)}{\rm{R}_f\rm{R}_b(\theta)},
\end{equation}

where $\rm{D}_f\rm{D}_b$, $\rm{D}_f\rm{R}_b$, $\rm{D}_b\rm{R}_f$ and $\rm{R}_f\rm{R}_b$ represent the normalised foreground-background, foreground-random, background-random and random-random pair counts at an angular separation $\theta$. For this work, the angular range was set from $\theta=10^{-0.3}$ arcmin (to ensure weak lensing regime) to $\theta=10^{2.1}$ arcmin, to avoid the large uncertainties of the large-scale datapoints.

The primary novelty of this work lies in the method used to generate the foreground and background random samples. Although \textit{Herschel} provided an instrumental noise map to account for the effect of the scanning strategy on overlapping zones – allowing for the inclusion of this effect in the generation of random catalogues for the H-ATLAS dataset – as previously mentioned, there is insufficient information to directly correct for these or similar effects in the other catalogues. As has already been pointed out, there are no additional science-ready catalogues of SMGs for cosmological purposes, since the submillimetre sky remains largely unexplored. Extracting cosmological information from these datasets therefore required the development of a new methodology to deal with such limitations. Hence, the random catalogues used in this work were generated using a Kernel Density Estimaton (KDE) technique. This approach was chosen not only to mimic the effect of the scanning strategy in surveys like H-ATLAS, but more importantly to address the significant spatial inhomogeneity present in catalogues such as HSPSC, which is not a systematic survey and therefore lacks coverage or selection criteria. The KDE-based method allows for the construction of random samples that are not purely random, but instead reflect — to some extent — the non-physical, systematic structure underlying the distribution of detections. These systematics can arise from instrumental and observational factors, such as varying depth, coverage, or scanning overlaps. 

\subsubsection{Generation of the random pairs: KDE}
\label{subsubsec:KDE}

KDE \citep{Sil86} is a non-parametric method used to estimate the continuous probability density function of a discrete dataset. It operates by placing a kernel function \( K \) with a specific shape and bandwidth \( b > 0 \) over each data point \( X_i \), producing a smooth approximation of the underlying distribution by summing these kernels across all data points.

The density estimate at a given point \( x \) is defined as:  
\begin{equation}
    \label{eq:KDE}
    \hat{f}(x;b) = \frac{1}{n b} \sum_{i=1}^{n} K\left(\frac{x - X_i}{b}\right),
\end{equation}  
where \( n \) is the number of data points, \( b \) controls the smoothness of the estimate, and \( K(\cdot) \) is a kernel function that integrates to one and typically satisfies smoothness and symmetry properties.

In this study, a Gaussian kernel was chosen due to its widespread use and robustness. This preference is largely because the kernel shape has a relatively minor impact on the density estimation compared to the bandwidth's influence.

While KDE is widely used and extensively studied, it has limitations with long-tailed distributions. Specifically, while a small bandwidth \( b \) can result in spurious noise in the tails of the estimates, increasing it to reduce this noise may obscure essential details in the central regions of the distribution. 

From a physical point of view, this variance-vs-bias trade-off affects the CCF signal as follows. If $b$ is too small (particularly if it drops below 0.1), then the estimator overfits the data distribution, producing a random catalogue that resembles the original data too much (imprinting the actual lensing signal), causing the lensing CCF signal to drop even at smaller scales where the signal is stronger. On the other hand, if $b$ is too big, the random catalogue will be too smooth: it will smooth out any clustering inherent to the data (both physical and systematic, as that derived from the scanning strategy), and it will produce a spurious CCF signal, particularly at medium and large angular distances where the real CCF signal is fainter. This latter case obviously applies only up to a given $b$ value, since for sufficiently large $b$ (approaching infinity), KDE would remove all structure completely and would produce purely random catalogues.


Bandwidth selection is critical, as it significantly affects the smoothness of the density estimate. Various methods have been proposed to determine the optimal bandwidth, most of which are based on minimising the Mean Integrated Squared Error (MISE)\footnote{For a detailed overview of bandwidth optimisation techniques, see \citet{Jin21}.}. 

For our purposes, a customised bandwidth selection method was developed to address specific characteristics observed in the data while preserving the lensing signal. It was hypothesised that inhomogeneities in the catalogue construction might cause small anomalies, such as bumps in the medium angular distance range, which cannot be explained by theoretical models. To mitigate this, a global bandwidth ($b$) was selected for each zone within each catalogue, guided by the following criteria:

\begin{enumerate}

\item Values of $b$ that caused the signal to decrease at small angular scales (typically below 1 arcmin) were discarded.

\item Values of $b$ that caused an increase at large angular scales were also discarded.

\item The optimal bandwidth was then selected from the remaining $b$ candidates by minimising the squared error between the cross-correlation obtained with each $b$ value and that obtained using a homogeneous random distribution. To compute this error, the most stable angular separation range—defined as the interval with the smallest error bars, typically between $\log(\theta [arcmin]) = -0.4$ and $\log(\theta [arcmin]) = 1.1$ for the current datasets—was used.

\end{enumerate}

Note that the $b$ value is the same for the foreground and background catalogue in each zone due to the optimising procedure. 

This approach ensured that KDE did not artificially reduce signals at small angular scales or introduce spurious signals at larger scales, thus in keeping with the rationale that these systematic effects are mainly concentrated on large scales.

By tailoring the bandwidth selection to the specific needs of this study, we aimed to achieve more accurate random catalogues that reflect the underlying data characteristics without introducing bias from catalogue inconsistencies. It is worth noting, however, that for most samples the optimal bandwidth was found to be slightly larger—by a factor of approximately 1.2—than that obtained using the standard automated leave-one-out cross-validation approach \citep[a widely used optimiser; see][]{HAS09}.

Once the optimal bandwidth was selected, the CCF was computed as the mean of 100 different realisations of the generated random samples. The bandwidth for each realisation was drawn from a Gaussian distribution with a mean equal to the optimal bandwidth and a standard deviation of 0.04.

While the use of the KDE technique is a powerful tool to address systematics arising from catalogue construction issues, it is important to note that other effects of physical origin (such as the presence of superclusters) will also impact the estimation of the angular CCF function and will be smoothed out by the KDE. Even if we do not yet have a theoretical model for these effects, they contain important information on large-scale structure that could be lost after smoothing. However, for the datasets used in this work, this is subdominant with respect to the systematics inherent to the catalogue

\subsubsection{Finite volume corrections: integral constraint}
\label{subsubsec:IC}

When working with small (finite) survey fields, a deviation arises between the expected and observed CCF, introducing a systematic negative offset. This effect occurs because the density fluctuations in the sample are normalized to the mean galaxy density within the observed region. In sufficiently large surveys, the mean galaxy density closely approximates the cosmic mean, as large-scale structure fluctuations are properly accounted for. However, in smaller fields (such as Cosmos and Bootes in HSPSC), these large-scale fluctuations are not fully sampled, leading to a local overdensity that overestimates the mean galaxy density. As a result, the estimated CCF is artificially suppressed \citep{Inf94, ROC99, ADE05}.

The true CCF, $w_{\text{real}}(\theta) $, and the observed one, $w_{\text{obs}}(\theta)$, are related by: $ w_{\text{obs}}(\theta) = w_{\text{real}}(\theta) - IC $, where IC is the offset known as the integral constraint. Numerically, this offset can be estimated by assuming that the true CCF follows a power law of the form $w_{\text{real}}(\theta) = A \theta^{-\alpha}$. Then, the IC is computed as:

\begin{equation}
    \text{IC} = \frac{\sum_i A \theta^{-\alpha} RR_i }{\sum_i RR_i}
\end{equation}

To determine IC, the observed CCF was first fitted to obtain the parameters \( A \) and \( \alpha \), and then these values were used to compute the IC for each field. An iterative approach was then implemented until convergence was reached, with a relative tolerance of $10^{-4}$. Nonetheless, the impact of the IC correction is small, as the majority of the regions are sufficiently large. 

\subsubsection{Error estimation}
\label{subsubsec:errors}

The covariance matrix was estimated using the Bootstrap resampling technique, following the methodology detailed in \citet{GON23} and \citet{CUE24}. The survey area was divided into $N$ equal-area sub-regions using a k-means algorithm to preserve spatial correlations. Bootstrap resampling was performed by selecting $N_r=3N$ patches with replacement and repeating the process $N_b=10000$ times, as recommended by \cite{NOR09} and validated in \cite{GON23}. This approach ensures a robust covariance estimation.

The covariance matrix is computed as

\begin{equation} \text{Cov}_{BS}(\theta_i,\theta_j)=\frac{1}{N_b-1}\sum_{k=1}^{N_b},\bigg[\hat{w}_k(\theta_i)-\bar{\hat{w}}(\theta_i)\bigg]\bigg[\hat{w}_k(\theta_j)-\bar{\hat{w}}(\theta_j)\bigg]\label{covariance}, \end{equation}

where $\hat{w}_k$ is the CCF function from the $k^{\text{th}}$ Bootstrap sample, and $\bar{\hat{w}}$ is its mean across all samples.

As explained in \ref{subsubsec:KDE}, the CCF was measured via the average over 100 realizations of the random catalogs. The errors associated with this technique, $\sigma_{\text{KDE}}$, were then added to the diagonal of the previously computed covariance matrix, yielding the final matrix: $\text{Cov}(\theta_i,\theta_j) = \text{Cov}_{BS}(\theta_i,\theta_j) + \delta_{ij} \sigma_{\text{KDE}}^2,$ where $\delta_{ij}$ is the Kronecker delta.

Finally, in this work we introduced a regularisation scheme for the covariance matrix. We observed that some of our covariance matrices are poorly conditioned, with condition numbers exceeding $10^5$–$10^6$. In such cases, matrix inversion becomes numerically unstable: even small amounts of noise in the covariance can produce large fluctuations in its inverse. Since our likelihood estimator explicitly depends on the inverse of the covariance matrix, a poor conditioning can introduce biases, underestimate parameter uncertainties, and generally amplify the estimation error \citep{LED04, TAB20, EUCLID25}. This issue becomes more severe when performing tomographic or multi-catalogue analyses, where convergence can be significantly hindered.

To address this, we applied a regularisation procedure prior to inverting the matrix, based on Tikhonov regularisation \citep{TIK63}. Specifically, we followed the methodology proposed by \citet{MAT12}, who define a regularised covariance matrix as:

\begin{equation}
\text{Cov}_{\mathrm{Reg}}(\theta_i, \theta_j) = \text{Cov}(\theta_i, \theta_j) + f \times \mathrm{Me}[\text{Cov}(\theta_i,\theta_j)],
\end{equation}

where $\mathrm{Me}[\text{Cov}(\theta_i,\theta_j)]$ denotes the median of the diagonal elements of the original (unregularised) covariance matrix, and $f$ is a regularisation factor that controls the level of conditioning. \citet{MAT12} found optimal values of $f$ to lie in the range of a few percent (typically 3–3.5\%).

The choice of $f$ reflects a compromise: larger values improve the stability of the inversion and suppress noise in the off-diagonal elements, but also widen the constraints by effectively increasing the variances and reducing the amount of cross-correlation information retained. In this work, we explored several values of $f$ (e.g. 0.0004, 0.001, 0.01, 0.026, 0.035, 0.1). To assess the impact of regularisation, we constructed mock cross-correlation measurements using the covariance structure of our datasets and evaluated the recovery of known (input) parameters. The choice $f = 0.035$ was found to improve convergence and slightly outperformed other values in terms of accuracy, yielding posterior distributions that remained broadly consistent—typically within $0.5\sigma$—with those obtained using the unregularised covariance matrix.

Although a more systematic optimisation of the regularisation factor $f$ is certainly possible, it lies beyond the scope of this work, given the choice of $f=3.5\%$ provides a practical and robust solution that ensures numerical stability without significantly altering the scientific conclusions.


\subsection{Parameter estimation}
\label{subsec:paramest} 

The theoretical model for the CCF used in this study, which is described in Appendix \ref{app:framework}, depends on a set of free parameters describing cosmology and the Halo Occupation Distribution (HOD) model. These parameters were explored using the MCMC method, implemented via the open-source package \texttt{emcee} \citep{FOR13}, based on the Affine Invariant MCMC Ensemble sampler \citep{GOO10}. The goal is to sample the posterior probability density function to obtain marginalised credible intervals and probability contours. The MCMC sampling used 21 walkers (3 times 7 parameters) and 15000 iterations. A burn-in phase was applied to remove initial iterations where the chains had not yet converged.

The measurement-based and theoretical CCF are computed as in equations \ref{eq:w_fb_obs} and \ref{eq:w_fb}. The log-likelihood function is given by:

\begin{align*}
    \log{\mathcal{L}\,(\theta_1,\ldots,\theta_m)}=-\frac{1}{2}&\bigg[m\log{(2\pi)}+\log{|C|}+\overrightarrow{\varepsilon}^{\text{T}}C^{-1}\,\overrightarrow{\varepsilon}\bigg],
\end{align*}
where 
$\overrightarrow{\epsilon}\equiv [\varepsilon(\theta_1),\ldots,\varepsilon(\theta_m)]$, 

\begin{equation*}
    \varepsilon(\theta_i)\equiv w(\theta_i)-\hat{w}(\theta_i)\quad\quad \forall i\in\{1,\ldots,m\},
\end{equation*}

and $C$ is the covariance matrix of the CCF measurements, as defined in eq. \eqref{covariance}.

Seven parameters were analysed, divided into astrophysical and cosmological categories. The astrophysical parameters, defining the HOD model \citep[see also Appendix \ref{app:framework}]{Cue23}, are $M_{min}$, $M_1$, and $\alpha$. The parameter $\beta$, related to the logarithmic slope of background source number counts, was also included. The cosmological parameters are $\Omega_m$, $\sigma_8$, and $h$, where $h$ is defined as $H_0=100\cdot h\text{ km}\text{s}^{-1}\text{Mpc}^{-1}$. A flat $\Lambda$CDM universe is assumed with $\Omega_{\Lambda}=1-\Omega_m$, while $\Omega_b$ and $n_s$ are fixed to their best-fit values, $\Omega_b=0.0486$ and $n_s=0.9667$\citep{PLA18_VI}. In line with the weak lensing approximation, only CCF data at angular scales $\geq 0.5$ arcmin are considered \citep{BON19}.
The choice of priors for each parameter is illustrated in Table \ref{tab:priors}. Uniform priors were adopted for both the astrophysical and cosmological parameters, with the latter being the same as those used in \cite{BON20, BON21}. Additionally, a Gaussian prior was applied to $\beta$, with the same values as in \citet{Cue23}.

\begin{table}[ht]
  \caption{Prior distributions for the MCMC analyses. }
  \label{tab:priors}
  \centering
  \begin{tabular}{cccc} 
    \hline\hline
    \multicolumn{2}{c}{Astrophysical} & \multicolumn{2}{c}{Cosmological} \\
    \cmidrule(r){1-2} \cmidrule(r){3-4}
    Parameter & Prior & Parameter & Prior\\
    \midrule
    $\log{M_{min}}$ & $\mathcal{U}[10.0-16.0]$
        & $\Omega_m$ & $\mathcal{U}[0.1-0.8]$\\
    $\log{M_{1}}$ & $\mathcal{U}[10.0-16.0]$
        & $\sigma_8$ & $\mathcal{U}[0.6-1.2]$\\
     $\alpha$ & $\mathcal{U}[0.5-1.5]$
        & $h$ & $\mathcal{U}[0.5-1.0]$\\
     \midrule
     $\beta$ & $\mathcal{N}[2.9 , 0.04]$\\
     \hline\hline
  \end{tabular}
  \tablefoot{Uniform priors, denoted as $\mathcal{U}$, are defined by their minimum and maximum values, while Gaussian priors, $\mathcal{N}$, are characterised by their mean ($\mu$) and standard deviation ($\sigma$).}
\end{table}


In addition to the individual MCMC analyses conducted for each dataset, joint analyses were also performed. The sky zones used in the joint analysis were chosen to avoid overlap, so they can be regarded as independent datasets. Consequently, a single model for MCMC with an extended parameter space was built, and then it was constrained by maximising a log-likelihood given by the sum of the log-likelihood functions of the different datasets. The number of parameters in each analysis included the three cosmological parameters ($\Omega_m, \sigma_8, h$), $\beta$, and $n \times 3$ HOD parameters ($\log M_1, \log M_{\text{min}}, \alpha$), where $n$ is the number of catalogues included in the analysis. The number of walkers was also scaled as three times the number of parameters in each case.

\section{Results}
\label{sec:results}

In this section the results of the CCF analysis and parameter inference for each new region will be presented and discussed. Then, parameter constraints derived from each individual catalogue will be examined. Finally, a joint analysis incorporating the entire explored area will be performed. In each section, the CCF analysis and MCMC results will be shown for the same regions as in \citet{CUE24} for comparison.

\subsection{Cross-correlation}
\label{res:xcorr}
\subsubsection{Methodology validation on GAMA/H-ATLAS data}

\begin{figure}[ht]
    \includegraphics[width=0.5\textwidth]{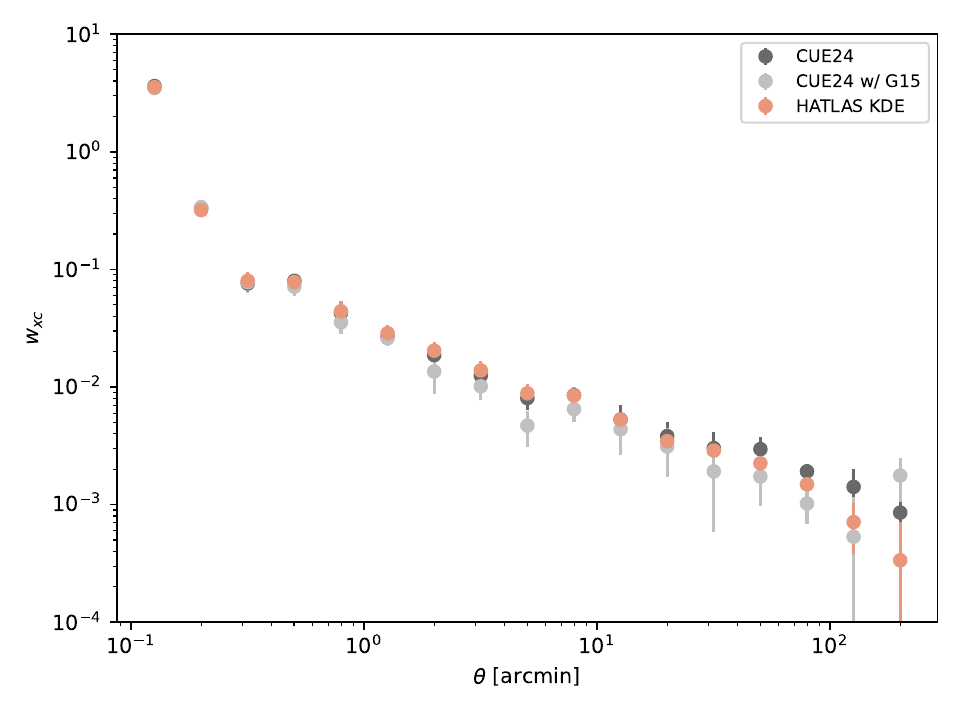}\\
     \caption{CCF measurements using GAMA survey as lenses and H-ATLAS SMGs as background sources. The data obtained with the new methodology that uses KDE to reproduce the random catalogue is represented in orange. Results from CUE24, where no such treatment was applied, are depicted in dark grey (CCF using the sky zones G09, G12, G15 and SGP) and light grey (CCF eliminating G15 from the analysis). 
     }
     \label{fig:xcorr_hatlas}
\end{figure}

The CCF on GAMA/H-ATLAS data was already studied in \citet{CUE24}. They found that one of the sky regions (G15) produced a signal excess at large angular scales, which was attributed to sampling variance. When this region was excluded, the signal amplitude decreased and the inferred cosmological parameters—particularly $\Omega_m$—shifted towards more standard values. Fig.~\ref{fig:xcorr_hatlas} presents the aforementioned CCF measurements obtained using the GAMA survey as lenses and the overlapping H-ATLAS SMGs as background sources. Their results using the full dataset are shown in grey. A clear excess is visible at angular separations beyond 60 arcmin, which \citet{CUE24} attributed to an excess of pairs in the G15 region. The measurements after removing G15 are shown in light grey.

As a proof of concept, we tested whether applying KDE to generate the random catalogues for the four GAMA regions could reproduce the effect of removing G15. The results obtained using the new methodology described in Section~\ref{sec:method}, applied to all four regions, are shown in orange in Fig.~\ref{fig:xcorr_hatlas}. At small angular separations, the measurements are in excellent agreement with those from \citet{CUE24} (dark grey, hereafter refer to as CUE24), while at larger scales they closely match the results obtained when G15 is excluded. This consistency suggests that proposed methodology can, to a certain extent, account for an effect like sampling variance on data pairs.

\subsubsection{Cross-correlation analysis on the new sky zones}

\begin{figure*}[h]
    \centering
    \includegraphics[width=1\textwidth]{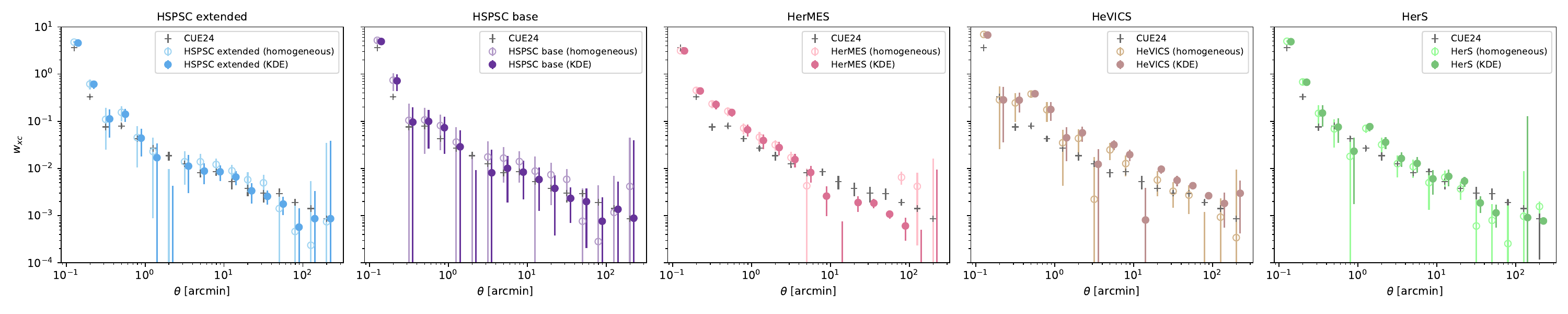}\\
     \caption{CCF estimated using a homogeneous random catalogue (older methodology, shown as empty circles) and a KDE-based random catalogue (filled circles), for the different datasets under study. The lenses are SDSS low-redshift galaxies, and the background sources are SMGs. In the first and second panels, HSPSC data are shown in blue (all available regions) and in purple (the “base” dataset, restricted to regions exclusive to HSPSC). The third panel displays HerMES data in pink, the fourth HeVICS in brown, and the fifth HerS in green. For comparison, results from CUE24 are overlaid in each panel as grey crosses. An offset has been applied to the KDE-based results along the x-axis for clarity.  
     }
     \label{fig:xcorr_panel}
\end{figure*}

In this subsection, we present CCF measurements obtained using the catalogues described in Section \ref{subsec:data}. Fig. \ref{fig:xcorr_panel} shows the CCF data obtained for each catalogue in this study. For each panel, results are shown using two different methods: a homogeneous random distribution (represented by empty circles) and a KDE-based random sample (filled circles). All the new measurements are corrected by the IC. The colour code used (and maintained throughout the paper) is as follows: HSPSC appears in blue for the extended dataset, which includes all available regions, and in purple for the base dataset, restricted to regions exclusive to HSPSC; HerMES is represented in pink; HeVICS in brown; and HerS in green. In every panel, the CCF data from the CUE24 dataset (GAMA/H-ATLAS) is overlaid as grey crosses for comparison.

The discussion begins with the results obtained using homogeneous randoms across all datasets. In the HSPSC datasets (first two panels of Fig. \ref{fig:xcorr_panel}), magnification bias is clearly observed, with low and mid angular distances following a trend very similar to CUE24. Large-scale points are less constrained, which can be attributed to the lower number of background sources: there were around 37000 well-detected SMGs in H-ATLAS (in an area of 207~deg$^2$) versus 12000 (in a similar area of 239~deg$^2$) in the HSPSC extended dataset. The number of lens candidates in each area is also different: the GAMA survey contains around 130000 galaxies in the overlapping area, while spectroscopic SDSS only 66700. Larger error bars in both extended and base datasets (compared to CUE24) can also be explained by poorer statistics, with the HSPSC base dataset showing the largest uncertainties among the three. Finally, a small signal excess appears in the intermediate angular range (between 4 and 40 arcmin). This excess is likely due to catalogue construction issue stemming from the fact that the catalogue was not originally designed for spatial correlation analyses. An illustrative example of such issues is shown and discussed in Fig. \ref{fig:zone-examples} in Appendix \ref{app:skyzones}. We hypothesise that this effect arises from the strong dependence of the signal on the sky area selection. When the selected regions include voids or galaxy overdensities—mainly caused by the scanning strategy—and no corresponding systematics masks are available, an artificial enhancement of the signal is observed at intermediate and large angular scales. 

The third panel in Fig. \ref{fig:xcorr_panel} shows the CCF obtained with the HerMES catalogue (pink empty circles). The results beyond 3 arcmin show a sharp drop where the signal becomes undetectable and loses all statistical significance. While HerMES consists of three different sky regions of a decent size, all of them were found to feature extreme spatial inhomogeneities in the spectroscopic lens catalogue. Indeed, while the HerMES SMG background sample is well behaved in this sense, spectroscopic SDSS lenses that were selected in these sky zones seem to mix targets from different programs, imprinting a spurious pattern in the sky that is not physical. An example of this behaviour is shown in the middle panel of Fig. \ref{fig:zone-examples}. 

Finally, the fourth and fifth panels in Fig. \ref{fig:xcorr_panel} depict the CCF measurements for HeVICS (pink) and HerS (green). Both catalogues consist of a single sky zone, which explains the larger error bars and the deviation from the expected behaviour — likely a consequence of the increased statistical variance from using only one small zone. Interestingly, the CCF function obtained for HeVICS is markedly strong at low angular distances. Physically, larger values in the 1-halo regime can be explained either by a steeper luminosity function (information enclosed in the $\beta$ parameter; see equation \ref{eq:w_fb} in Appendix \ref{app:framework}) or by a larger mass of the lenses. All the galaxies used as sources in this work are submillimetre, and a $\beta$ value around 3 has been assumed for all of them, according to the literature \citep[for a detailed discussion, see][]{CUE24}. However, if the $\beta$ estimate is correct, the strength of the signal is comparable to that obtained using clusters of galaxies as lenses in previous studies \citep{FER24}.

Overall, regardless of the catalogue, apart from deviations that can be attributed to sample inhomogeneities, larger angular distances are mostly not significant (either no signal or very large error bars), which hinders the possibility of obtaining strong cosmological constraints. This is not a limitation of the technique itself, but of the size of the sky fields used. A new submillimetre survey could provide larger and more homogeneous zones and contribute to recovering the signal at larger angular scales.

The CCFs obtained for each catalogue using KDE-generated randoms are shown as filled circles in Fig. \ref{fig:xcorr_panel}. For the HSPSC datasets (blue and purple), while low angular distances are only slightly affected, the mid and large angular distances (particularly the 2-halo regime) align better with the expected behaviour of a cross-correlation: a steady signal decrease with no irregularities. The mid angular scales are in perfect agreement with CUE24, and the large angular scales behave as expected. Overall, the CCF obtained using KDE randoms resembles better the results obtained with similar lenses and background galaxies.

The HerMES dataset (pink, middle panel) shows that the CCF signal is recovered for mid and large angular distances compared to the homogeneous case. Finally, the cross-correlation of the HeVICS and HerS datasets also show slight improvements (particularly at large scales, where the signal is recovered), even though the HeVICS signal remains anomalously strong, and both still display irregularities, likely caused, as discussed, by the fact that they consist of a small single sky zone.

\subsection{MCMC Analysis}

\subsubsection{Methodology validation on GAMA/H-ATLAS data}

\begin{figure}[h]
    \includegraphics[width=0.5\textwidth]{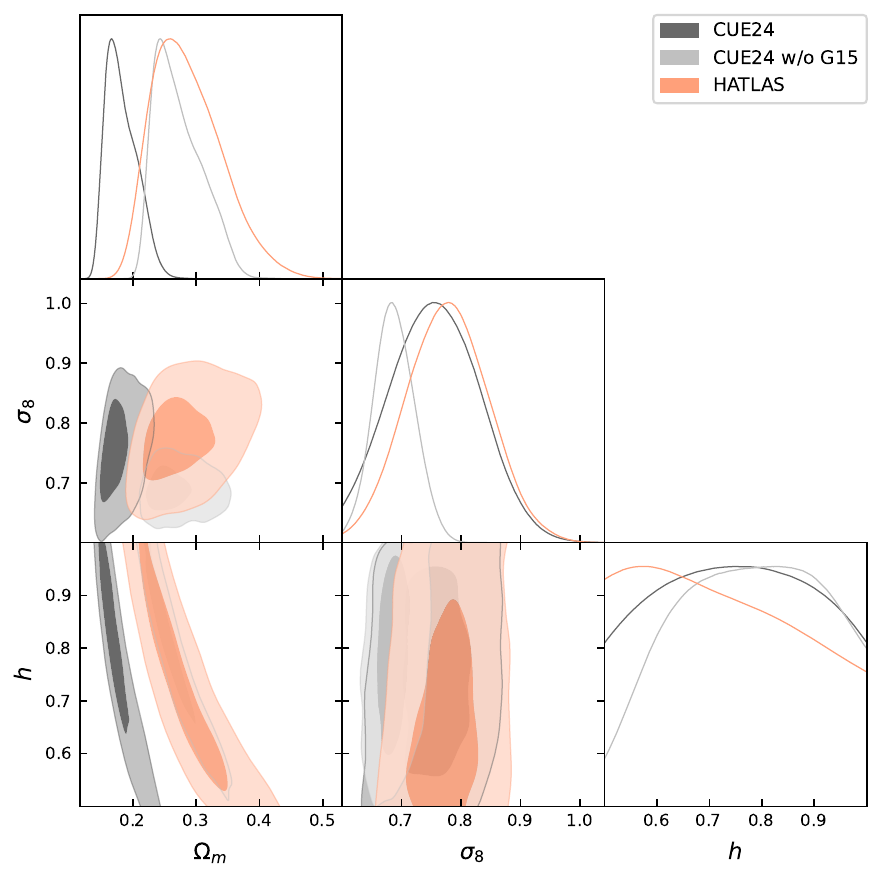}\\
     \caption{Marginalised posterior distributions and probability contours for the cosmological parameters in the in the GAMA/H-ATLAS case. The CCF obtained using KDE for the random catalogue is shown in orange, while the results from CUE24 are depicted in dark grey (including all four sky regions: G09, G12, G15, and SGP) and in light grey (excluding G15 from the analysis).
     }
     \label{fig:hatlas_cosmo}
\end{figure}

\begin{figure*}[h]
    \centering

    \begin{subfigure}[b]{0.32\textwidth}
        \includegraphics[width=\textwidth]{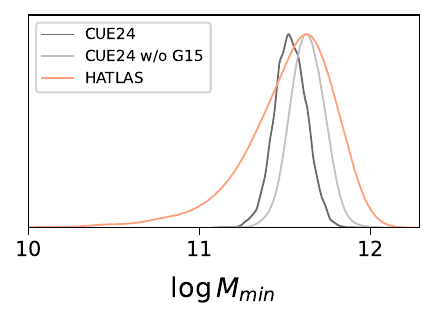}
    \end{subfigure}
    \begin{subfigure}[b]{0.32\textwidth}
        \includegraphics[width=\textwidth]{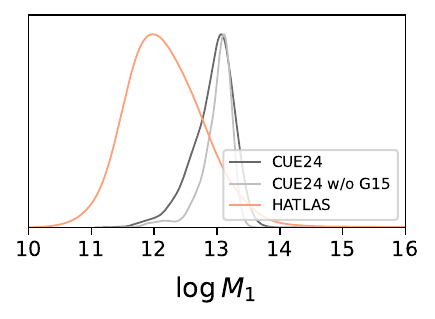}
    \end{subfigure}
    \begin{subfigure}[b]{0.32\textwidth}
        \includegraphics[width=\textwidth]{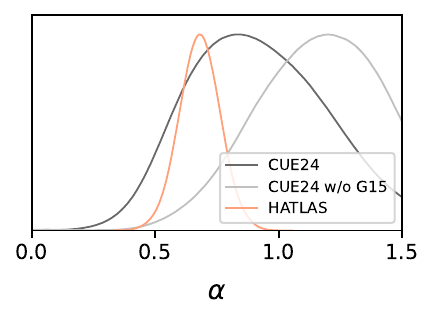}
    \end{subfigure}

    \vspace{0.5em} 

    \begin{subfigure}[b]{0.32\textwidth}
        \includegraphics[width=\textwidth]{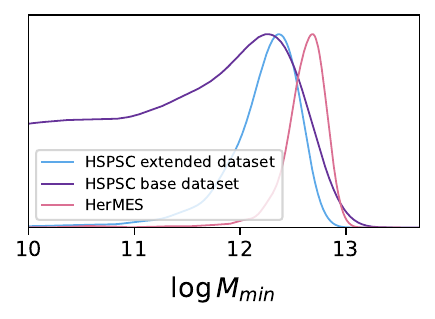}
    \end{subfigure}
    \begin{subfigure}[b]{0.32\textwidth}
        \includegraphics[width=\textwidth]{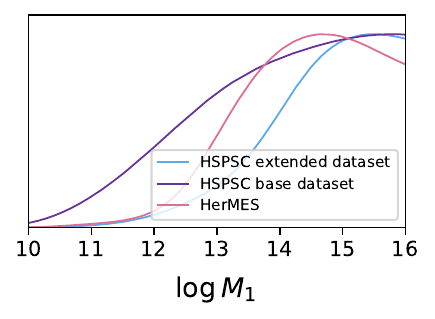}
    \end{subfigure}
    \begin{subfigure}[b]{0.32\textwidth}
        \includegraphics[width=\textwidth]{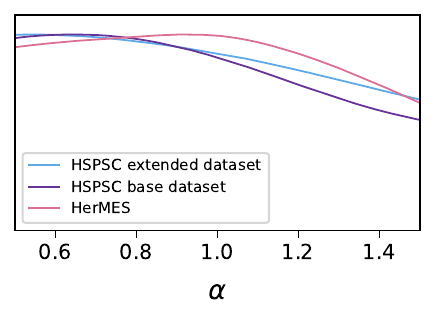}
        
    \end{subfigure}

    \caption{Marginalised posterior distribution of the HOD parameters derived from MCMC runs. Upper panel: GAMA and H-ATLAS data. The results obtained with the new methodology, which employs KDE to generate the random catalogue, are shown in orange. In contrast, the results from CUE24, where a homogeneous random catalogue was used, are depicted in dark grey (for the CCF including all sky regions: G09, G12, G15, and SGP) and in light grey (for the CCF excluding G15 from the analysis). Bottom panel: Marginalised posterior distribution of the HOD parameters derived from MCMC runs on the cross-correlation using SDSS as lenses and HSPSC extended (light blue), base (violet), and HerMES (pink) catalogues as sources.}
    \label{fig:hod}
\end{figure*}

Fig. \ref{fig:hatlas_cosmo} shows the marginalised posterior distributions and probability contours for the for the GAMA/H-ATLAS cross-correlation data. As before, our results are depicted in orange, while the two cases discussed in \citet{CUE24} are shown in dark grey (including all four sky regions) and light grey (excluding G15). The same colour scheme will be used throughout this section.

The results show great consistency with the results by CUE24 within the framework of KDE. In particular, for $\Omega_m$, the posterior distribution is closely matched to those of CUE24 when the G15 region (responsible for the anomalous large-scale excess) is excluded. This is to be expected, since cosmological parameters are strongly determined by the two-halo region, which lies at larger angular distances. More specifically, the obtained constraints (which are presented in Table \ref{tab:hatlas_results}) are as follows, with all upper and lower limits corresponding to the 68\% confidence interval.
$\Omega_m = 0.29^{+0.04}_{-0.07}$ using the new methodology, compared to $0.18^{+0.02}_{-0.03}$ for CUE24 and $0.27^{+0.03}_{-0.05}$ for CUE24 excluding G15; and $\sigma_8=0.78^{+0.07}_{-0.07}$ for the new methodology, compared to $0.76^{+0.07}_{-0.08}$ and $0.69^{+0.03}_{-0.04}$ for CUE24 and CUE24 excluding G15, respectively. While $\sigma_8$ is in good agreement between both methodologies, the peak for $\Omega_m$ is slightly closer to the 2018 \textit{Planck} value ($\Omega_m=0.315\pm 0.007$) than the other two cases. The constraints on $\Omega_m$ rely heavily on the large-scale points. During the analysis, it was observed that small variations in the final points used for the MCMC significantly impact this parameter. Although the results remain consistent with previous analyses, the deviation in the $\Omega_m$ peak is attributed to this effect. This dependency further underscores the need for larger survey areas in magnification bias studies. Expanding the sky coverage would help mitigate this strong dependence by averaging over different sky regions. As for \textit{h} (which remains a loosely constrained parameter in lensing studies), only rough constraints were obtained in CUE24 when excluding G15: $h=0.77^{+0.17}_{-0.12}$, while the parameter remained unconstrained when using all available area for the study. The upper limit obtained with the new methodology is $h<0.81$ (with a tentative peak on $h=0.73$), which is in perfect agreement with the previous analysis. 

Regarding the astrophysical parameters, the upper panel of Fig. \ref{fig:hod} shows the 1D marginalised posterior distributions for the three HOD parameters: $\log M_{\text{min}}$, $\log M_1$ and $\alpha$. The results from the analysis using the new treatment (depicted in orange) are in agreement with those of CUE24 including all four sky zones. This behaviour is expected, as the astrophysical parameters are primarily determined by the 1-halo regime, which corresponds to smaller angular scales, specifically left untouched by the choice of bandwidth. The most notable difference in the HOD constraints is found in $\log M_1$, whose posterior peak shifts from 13 in CUE24 to 12.2 with the new methodology. This deviation, at the $\sim$1$\sigma$ level, is likely the result of a combination of effects—such as subtle changes in the 1-halo regime due to the new random catalogue and minor modifications in the theoretical model used for the cross-correlation. However, the dominant factor is likely the degeneracy of $\log M_{\text{min}}$ with other parameters, both astrophysical and cosmological (notably with $\sigma_8$), as clearly seen in the full corner plot presented in Fig.\ref{fig:HATLAS-fullcorner}. Finally, a remarkable improvement is also observed in the precision of the $\alpha$ parameter estimation, whose uncertainty has been reduced by nearly a factor of four. In conclusion, the results obtained applying the new methodology using KDE to estimate the random are an improvement although consistent with the previous results using GAMA/H-ATLAS cross-correlation data.

\subsubsection{Parameter constraints on the new sky zones}

In this section, cosmological and astrophysical constraints on the new individual catalogues are presented. In particular, results are shown for the HSPSC extended dataset (light blue), the HSPSC base dataset (purple), and the HerMES dataset (pink). This colour scheme will be used consistently throughout the section. The HeVICS and HerS datasets were not used individually to constrain parameters due to the high variance arising from their limited sky coverage. 

\begin{figure}[h]
    \includegraphics[width=0.5\textwidth]{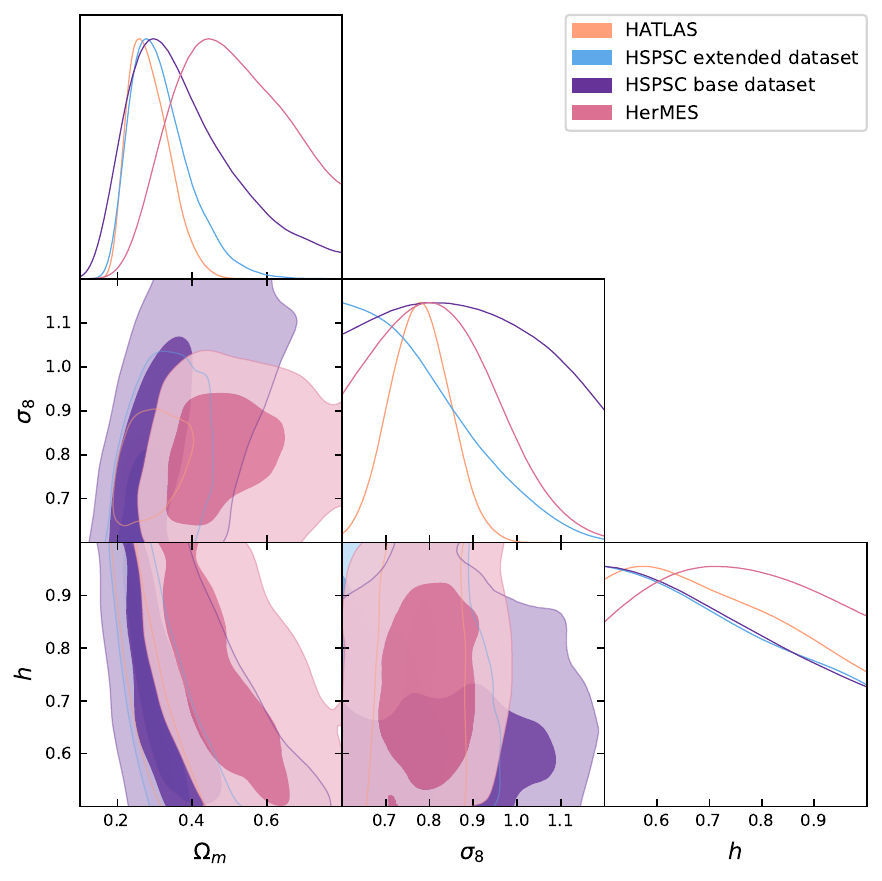}
     \caption{Marginalised posterior distributions and probability contours for the cosmological parameters in the HSPSC extended (light blue) and base (purple), and HerMES (pink) sources datasets. Overlapping areas of SDSS have been used as lenses in each case. GAMA/H-ATLAS results are depicted in orange for comparison.
     }
     \label{fig:newcats_cosmo}
\end{figure}

Fig. \ref{fig:newcats_cosmo} shows the marginalised posterior distribution for the cosmological parameters. Both $\Omega_m$ and $\sigma_8$ are less constrained than in the H-ATLAS study. The HSPSC extended and base datasets are in excellent agreement, with the base case showing slightly broader distributions, as expected. $\Omega_m$ is consistent with the \textit{Planck} 2018 results across both datasets, with values of $\Omega_m=0.32^{+0.05}_{-0.10}$ for the HSPSC extended dataset and $\Omega_m=0.39^{+0.08}_{-0.19}$ for the HSPSC base dataset. An anomalously high value of $\Omega_m=0.51^{+0.13}_{-0.17}$ was retrieved for the HerMES dataset. $\sigma_8$ results are also consistent across the three datasets: $\sigma_8<0.82$ with a tentative peak at $0.78$ and $\sigma_8=0.88^{+0.13}_{-0.25}$ for HSPSC extended and base respectively, and $\sigma_8=0.82^{+0.10}_{-0.17}$ for HerMES. Finally, only rough upper limits were obtained for \textit{h}: 0.80 for both HSPSC datasets and 1.00 for HerMES.

The discrepancies in the cosmological constraints from HerMES appear to stem from the way the covariance matrix is constructed using the sample data. As previously discussed, HerMES sky regions contain lens galaxy overdensities unrelated to physical structure. While the CCF can be partially corrected using KDE methods, the covariance matrix—derived directly from the data—is more affected. When the full angular range ($\log\theta = -0.3$ to $\log\theta = 2.1$) is used, MCMC fitting fails to reproduce the CCF, particularly above 7 arcmin, where it drops sharply or becomes negative. This scale coincides with the angular separation at which the CCF vanishes when using a homogeneous random catalogue (see Fig. \ref{fig:xcorr_panel}). Excluding the three angular bins between 7 and 20 arcmin (which include a negative CCF value and its surroundings) restores the fit, though with a slight underestimation. Mock catalogue tests show that even when using the theoretical CCF as input, the MCMC fails to recover a good fit if those problematic angular scales are included. In contrast, using the full angular range with a diagonal covariance matrix yields similar constraints to those obtained by excluding the affected bins.

Therefore, the results reported here exclude the three problematic angular scales, as their inclusion distorts the fit—even for ideal input data. A strictly rigorous solution would involve replacing the SDSS spectroscopic lens sample in the HerMES region, but we decided to keep it in this work to maintain a homogeneous treatment of the foreground sample for all the regions.

The HOD parameter marginalised posterior distributions are shown in the lower panel of Fig. \ref{fig:hod}. Results are consistent, but poorly constrained, across the three datasets. For $\log M_{\text{min}}$, the base dataset is slightly shifted towards lower values ($\log M_{\text{min}}=11.56^{+0.10}_{-0.60}$ compared to $\log M_{\text{min}}=12.20^{+0.31}_{-0.17}$ in the HSPSC extended dataset), although the peaks of the distributions are aligned. For HerMES, the retrieved value was $\log M_{\text{min}}=12.58^{+0.25}_{-0.10}$, closer to the extended dataset. On the other hand, only lower limits could be established for the $\log M_1$ parameter in both HSPSC datasets, at 14.31 and 13.31 for the extended and base samples, respectively. In HerMES, the value is lower, $\log M_1=11.57^{+0.73}_{-0.88}$. As for $\alpha$, an upper limit of 1.11 (mean 0.96) was obtained from the HSPSC extended dataset, while no constraints were retrieved from the base dataset. In HerMES, a lower limit of 1.13 was obtained.

Compared to the HOD results derived from the GAMA/H-ATLAS sample, the larger uncertainties in the CCF data led to broader posterior distributions compared to those derived from the GAMA/H-ATLAS sample. Nevertheless, the retrieved constraints are broadly consistent with the expected properties of SDSS galaxies \citep[see, e.g.][]{Tin05, Zeh05, Whi11, Zeh11, Guo15}, considering that the sample was not homogeneous in terms of galaxy types or redshift. In particular, the minimum mass is slightly higher than that of the GAMA galaxies (shifting from a mean value of 12–12.5 to nearly 13 in logarithmic scale). In this work, a single redshift bin was used for the foreground sample. Introducing redshift or luminosity binning could allow for a more detailed investigation of the HOD parameters and the dependence of clustering on halo mass and galaxy evolution. However, due to the limited statistics in these datasets, no further studies were carried out.
 
In summary, this analysis demonstrates that meaningful cosmological and astrophysical constraints can be derived from cross-correlation measurements in newly observed sky regions. While the parameter estimates are limited by the relatively low source density and sky coverage of the available datasets, the results validate the potential of submillimetre-selected samples for halo modelling and large-scale structure studies. 

\subsection{A joint analysis combining submillimetre catalogues}

Although SMGs have proven to be excellent sources for magnification bias studies, the limited sky coverage in this frequency range currently restricts their use as an independent cosmological tool. The results presented in the previous section required a significant homogenisation effort, yet they only yielded relatively broad constraints that, on their own, are not competitive with those obtained from the H-ATLAS catalogue. Nevertheless, they could still be combined with the H-ATLAS data into a joint analysis. Given that this work makes use of most of the available regions (a few Herschel-covered areas were excluded due to lack of overlap with SDSS), such a joint analysis would illustrate the current potential of SMGs for cosmic magnification studies over a single broad lens redshift bin.

\begin{figure}[ht]
    \includegraphics[width=0.5\textwidth]{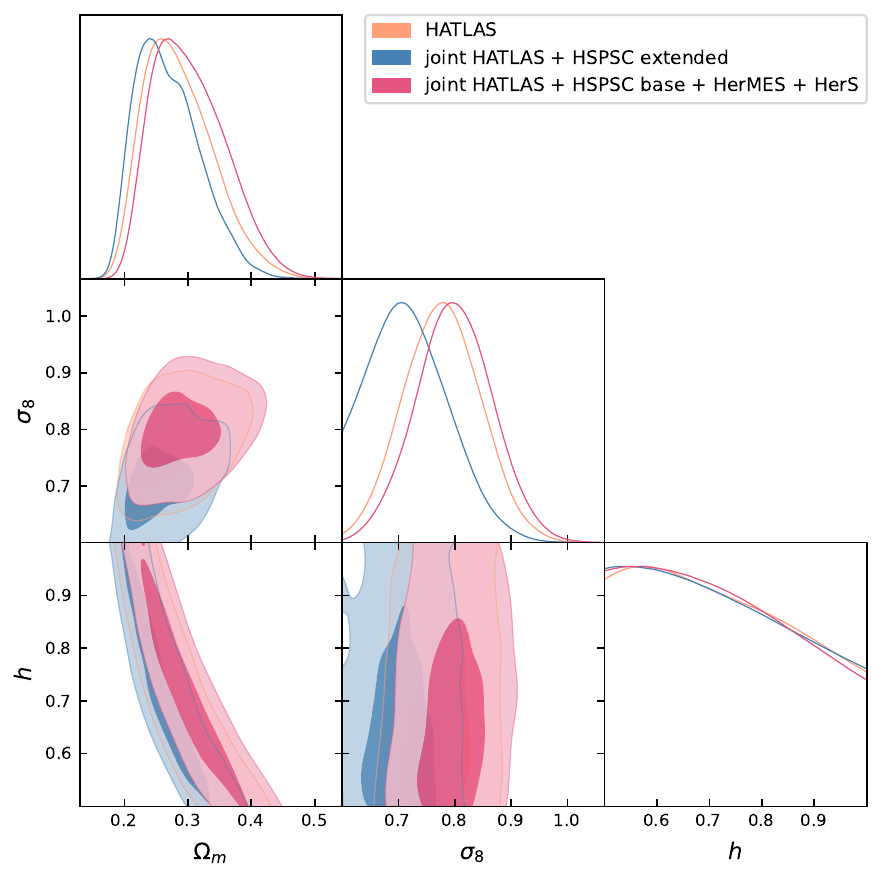}
     \caption{Marginalised posterior distributions and probability contours for the cosmological parameters in the different joint analyses: H-ATLAS + HSPSC extended datasets (JOINTx2), shown in blue, and H-ATLAS + HSPSC base + HerMES + HerS datasets (JOINTx4), shown in pink. The results from the H-ATLAS dataset alone are shown in orange for comparison. 
     }
     \label{fig:joint_cosmo}
\end{figure}

Fig. \ref{fig:joint_cosmo} shows the marginalised posterior distributions for two different joint analyses: H-ATLAS + the HSPSC extended catalogue (in blue, refered to as JOINTx2 onwards), and H-ATLAS + HSPSC base catalogue + HerMES + HerS (in pink, refered to as JOINTx4). In the latter case, since HeVICS was excluded from the MCMC analysis due to its anomalously strong signal, that region was incorporated into the HSPSC. The results from H-ATLAS alone (using the KDE-generated background) are shown in orange for comparison. 

As expected, the results are largely driven by the H-ATLAS dataset, due to its much smaller CCF uncertainties. The $\Omega_m$ estimates are very consistent across the three analyses, with only a minor shift in the peak of the maximum likelihood. In particular, the constraints are $\Omega_m=0.27^{+0.04}_{-0.06}$ for JOINTx2, $\Omega_m=0.30^{+0.05}_{-0.07}$ for JOINTx4, and $\Omega_m=0.29^{+0.04}_{-0.07}$ for H-ATLAS alone. The parameter $\sigma_8$ shows a similar trend: only the JOINTx2 estimate is slightly lower than that from H-ATLAS alone. The recovered constraints are $\sigma_8=0.72^{+0.06}_{-0.09}$, $\sigma_8=0.80^{+0.07}_{-0.07}$, and $\sigma_8=0.78^{+0.07}_{-0.06}$ for JOINTx2, JOINTx4, and H-ATLAS, respectively. As per \textit{h}, the three analysis remain consistent: only an upper limit of $\sim 0.80$ (with a tentative peak at $\sim 0.72$) is obtained in all three cases. All the constraints are shown in Table \ref{tab:joint_results}. 

Interestingly, JOINTx4 aligns more closely with H-ATLAS than JOINTx2 does. This could be due to the greater influence of the full HSPSC catalogue in JOINTx2; given that JOINTx4 includes more sky regions divided in a greater number of catalogues (with different construction effects each), it is plausible that this leads to a better statistical balance.

Although the constraints are in perfect agreement with $\Lambda$CDM, there is another potential source of error, specific to magnification bias studies, that can affect them and could contribute deviations, particularly in the new catalogues. The impact of the $\beta$ parameter, particularly on $\sigma_8$, has already been discussed for the GAMA–H-ATLAS dataset \citep{CUE24}. In this work, the $\beta$ prior distribution from \cite{CUE24} was adopted for all datasets, given that the nature of the sources is the same. A precise determination of the $\beta$ parameter for the new catalogues and sky regions is not straightforward, due to completeness and other catalogue-construction issues, and was beyond the scope of this study. However, it is possible that the actual value of $\beta$ may lie slightly above or below the H-ATLAS value of 2.9. Given the broad constraints, this would likely have minimal impact on the general results, but could slightly shift the peak of the probability distribution for $\sigma_8$.

In conclusion, although the new catalogues—processed with the revised method for generating randoms—yield cosmological constraints individually, they are broader than those derived from H-ATLAS. However, a joint analysis using all available SMG datasets provides perfectly compatible results with respect to the H-ATLAS catalogue. 

\section{Conclusions}
\label{sec:concl}

Magnification bias has consistently demonstrated its value as an independent cosmological probe over the past decade. However, the selection of suitable background sources is critical in order to obtain a significant and stable CCF that can be used for inference. In this context, SMGs have proven to be an excellent choice. Nonetheless, the amount of sky surveyed in the submillimetre range remains limited, largely due to the lack of wide-area studies specifically designed for statistical analysis in this regime.

The aim of this work has been to assess the potential of magnification bias on SMGs by extending the GAMA/H-ATLAS regions using the sky surveyed by Herschel. To this end, source samples were built based on different catalogues (HSPSC, HerMES, HeVICS, HerS).

Unlike H-ATLAS, these new catalogues were not initially designed for statistical studies. Therefore, the randoms were generated using a novel method based on KDE techniques. This new approach enabled the measurement of the CCF in all the newly explored regions, yielding results consistent with the presence of magnification bias. Importantly, this constitutes the first measurement of magnification bias in SMGs beyond the H-ATLAS catalogue, an observable that has so far been relatively unexplored, and thus provides a valuable validation of its detectability. Furthermore, the new methodology provides a more robust detection of the CCF signal than the previous method, showing good agreement with cross-correlations obtained from SMGs in earlier studies. However, measurements on large angular scales still suffer from considerable uncertainties, likely due to the size and number density distribution of the surveyed areas, which limit the cosmological information that can be extracted. Finally, although magnification bias was consistently detected across all catalogues, both HeVICS and HerS—being single-field catalogues—exhibit anomalies probably linked to sampling variance, rendering them suboptimal for inference on their own. Despite the fact that the catalogues employed in this work were not optimised for statistical analyses, reasonable CCF measurements and parameter constraints were obtained for the new sky zones. However, none of the new datasets can individually match the performance of the GAMA/H-ATLAS combination analysed in \citet{CUE24}. This is unsurprising given their smaller number of background sources and lens candidates and, most importantly, the survey nature of the data.

Regardless of all the efforts, the joint analysis of the new catalogues alongside H-ATLAS shows a similar constraining power to that of using H-ATLAS alone, suggesting that the characteristics and sky coverage of the additional data are not well suited to improve the constraints in this context. There is still room for further analysis of the Herschel SMGs population, such as a tomographic redshift-dependent approach (as in \citet{BON24}) or a a photometric lens sample with a much higher number density. However, without new catalogues covering wider sky areas, the potential to improve the precision of current constraints remains limited.

\begin{acknowledgements}
RFF, LB, DC, JGN and JMC acknowledge the PID2021-125630NB-I00 project funded by MCIN/AEI/10.13039/501100011033/FEDER, UE.
LB, JGN and JMC also acknowledges the CNS2022-135748 project funded by MCIN/AEI/10.13039/501100011033 and by the EU “NextGenerationEU/PRTR”. \\
The \textit{Herschel}-ATLAS is a project with \textit{Herschel}, which is an ESA space observatory with science instruments provided by European-led Principal Investigator consortia and with important participation from NASA. The H-ATLAS web- site is http://www.h-atlas.org. GAMA is a joint European- Australasian project based around a spectroscopic campaign using the Anglo- Australian Telescope. The GAMA input catalogue is based on data taken from the Sloan Digital Sky Survey and the UKIRT Infrared Deep Sky Survey. Complementary imaging of the GAMA regions is being obtained by a number of independent survey programs including GALEX MIS, VST KIDS, VISTA VIKING, WISE, \textit{Herschel}-ATLAS, GMRT and ASKAP providing UV to radio coverage. GAMA is funded by the STFC (UK), the ARC (Australia), the AAO, and the participating institutions. The GAMA web- site is: http://www.gama-survey.org/.\\
Funding for the Sloan Digital Sky Survey V has been provided by the Alfred P. Sloan Foundation, the Heising-Simons Foundation, the National Science Foundation, and the Participating Institutions. SDSS acknowledges support and resources from the Center for High-Performance Computing at the University of Utah. SDSS telescopes are located at Apache Point Observatory, funded by the Astrophysical Research Consortium and operated by New Mexico State University, and at Las Campanas Observatory, operated by the Carnegie Institution for Science. The SDSS web site is \url{www.sdss.org}.

SDSS is managed by the Astrophysical Research Consortium for the Participating Institutions of the SDSS Collaboration, including Caltech, The Carnegie Institution for Science, Chilean National Time Allocation Committee (CNTAC) ratified researchers, The Flatiron Institute, the Gotham Participation Group, Harvard University, Heidelberg University, The Johns Hopkins University, L'Ecole polytechnique f\'{e}d\'{e}rale de Lausanne (EPFL), Leibniz-Institut f\"{u}r Astrophysik Potsdam (AIP), Max-Planck-Institut f\"{u}r Astronomie (MPIA Heidelberg), Max-Planck-Institut f\"{u}r Extraterrestrische Physik (MPE), Nanjing University, National Astronomical Observatories of China (NAOC), New Mexico State University, The Ohio State University, Pennsylvania State University, Smithsonian Astrophysical Observatory, Space Telescope Science Institute (STScI), the Stellar Astrophysics Participation Group, Universidad Nacional Aut\'{o}noma de M\'{e}xico, University of Arizona, University of Colorado Boulder, University of Illinois at Urbana-Champaign, University of Toronto, University of Utah, University of Virginia, Yale University, and Yunnan University.

This research has made use of the python packages \texttt{ipython} \citep{ipython}, \texttt{matplotlib} \citep{matplotlib} and \texttt{Scipy} \citep{scipy}.
\end{acknowledgements}

\bibliographystyle{aa} 
\bibliography{xc_clusters} 

\begin{appendix}

\section{Theoretical framework}
\label{app:framework}

Magnification bias of SMGs provides a valuable cosmological probe through the lensing-induced CCF between background SMGs and foreground galaxies \citep[e.g.,][]{GON17, BON20, CUE21, GON21}. This effect, widely studied in the literature \citep[see][]{Bar01}, arises from weak gravitational lensing that modifies both the observed flux and angular size of distant sources. When a flux limit is applied, these distortions lead to an apparent excess of background sources near foreground structures—commonly referred to as lenses—compared to unlensed regions.

 The theoretical CCF is computed under the Limber and flat-sky approximations as:

\begin{equation}
    \label{eq:w_fb}
    w_{\text{fb}}(\theta)=2(\beta -1)\int^{\infty}_0 \frac{dz}{\chi^2(z)}\frac{dN_f}{dz}W^{\text{lens}}(z)\int_{0}^{\infty}\frac{ldl}{2\pi}P_{\text{g-m}}(l/\chi^2(z),z)J_0(l\theta),
\end{equation}

where the lensing efficiency kernel $W^{\text{lens}}(z)$ is given by:

\begin{equation}
    W^{\text{lens}}(z)=\frac{3}{2}\frac{H_0^2}{c^2}E^2(z)\int_z^{\infty} dz' \frac{\chi(z)\chi(z'-z)}{\chi(z')}\frac{dN_b}{dz'},
\end{equation}

and $E(z)=\sqrt{\Omega_m(1+z)^3+\Omega_{\Lambda}}$. The terms $dN_f/dz$ and $dN_b/dz$ represent the normalised redshift distributions of the foreground and background populations, respectively, $\chi(z)$ is the comoving distance, and $P_{\text{g-m}}$ denotes the galaxy–matter cross-power spectrum. The slope $\beta$ corresponds to the logarithmic derivative of background number counts with respect to flux, and plays a key role in the amplitude of the signal \citep{Cue23}.

To alleviate the computational cost associated with the nested integrals in Eq.~\ref{eq:w_fb}, we adopt the mean-redshift approximation developed in \citet{Cue23}, which replaces the full redshift integral with an evaluation at the mean redshift $\bar{z}$. The approximate expression becomes:

\begin{equation}
    \label{eq:w_fb_approx}
    w_{\text{fb}}(\theta)\approx 2(\beta-1)\frac{W^{\text{lens}}(\bar{z})}{\chi^2(\bar{z})}\int_0^{\infty}\frac{ldl}{2\pi}P_{\text{g-m}}(l/\chi(\bar{z}),\bar{z})J_0(l\theta).
\end{equation}

The galaxy–matter cross-power spectrum $P_{\text{g-m}}$ represents the relationship between the matter density field and the galaxy distribution. Within the halo model \citep{COO02}, all matter is assumed to be bound within dark matter halos. On small angular scales, the signal is governed by correlations within single halos (the 1-halo term), while at larger scales, the clustering between distinct halos dominates (2-halo term).

The connection between galaxies and halos is specified through a Halo Occupation Distribution (HOD), which describes how galaxies populate halos of different masses. This work uses the three-parameter HOD model from \citet{ZHE05}, in which halos above a minimum mass $M_{\text{min}}$ host a central galaxy, and those above $M_1$ can also host satellites. The satellite population follows a power law in halo mass with slope $\alpha$. The occupation functions are defined as:

\begin{equation}
    \label{eq:ncen01}
    N_\text{c}(M) =
    \begin{cases}
    0, & \text{if } M < M_\text{min} \\
    1, & \text{otherwise}
    \end{cases}
\end{equation}

\begin{equation}
    \label{eq:nsat01}
    N_\text{s}(M) = N_\text{c}(M) \cdot \left(\frac{M}{M_1}\right)^{\alpha},
\end{equation}

with $M_\text{min}$, $M_1$, and $\alpha$ as free parameters. This parametrisation captures both the threshold for galaxy formation and the scaling of satellite abundance with halo mass.

\clearpage
\onecolumn
\section{Sky zones}
\label{app:skyzones}
The different sky zones analysed in this work are labelled in blue in Fig. \ref{app:skyzones}. The galaxies shown in the sky plot are sourced from the HSPSC catalogue, and minor variations in area coverage for the overlapping zones may occur across the different catalogues. Readers can refer to the respective catalogue publications for precise area coverage and data sources (SDSS: \citet{SDSSV}; HerMES: \citet{HerMES}; HeVICS: \citet{HeVICS}; and HerS: \citet{HerS}). The mapping of regions per catalogue is as follows: SGP, NGP, GAMA-9h, GAMA-12h and GAMA-15h in H-ATLAS; Bootes, ELAIS-N2, HeVICS, Hershel Stripe 82 (L and R), XMM-LSS, Lockman-SWIRE and COSMOS in HSPSC; and Lockman-SWIRE, XMM-LSS and ELAIS-N1 in HerMES. Both HeVICS and HerS are single-zone catalogues.    

\begin{figure*}[htpb]
  \centering
  \includegraphics[width=0.9\textwidth]{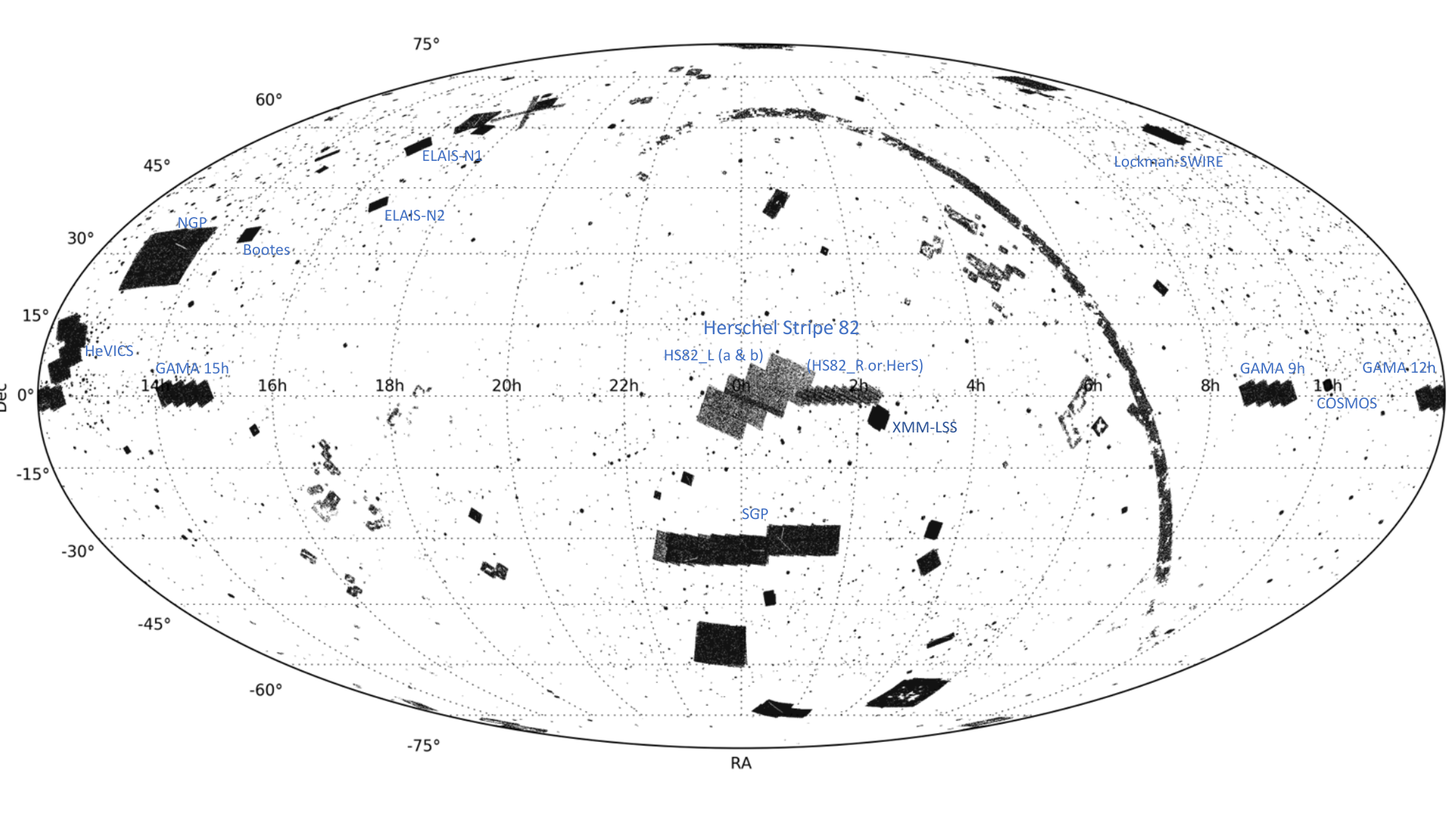}
  \caption{Sky view illustrating the coverage zones analysed in this study. Galaxies marked in black are sourced from the HSPSC catalogue, with minor variations in coverage area expected across the other catalogues. Readers seeking detailed plots can refer to the respective catalogue papers, though the depiction here provides an accurate overview for gaining a general understanding.}
  \label{fig:sky_zones}
\end{figure*}

As discussed in Section~\ref{subsec:data}, the lower quality of the new catalogues required a different selection methodology. The goal was to minimise inhomogeneities while retaining an area above 4 deg$^2$. While H-ATLAS showed a clear 30 mJy threshold from number counts, such a cut was less evident in the other catalogues, possibly due to incompleteness. The final selection applied this flux cut and required detection at 250 or 350 $\mu$m with S/N~$>4$.

\begin{figure*}[ht]
    \centering
    \includegraphics[width=0.15\textwidth]{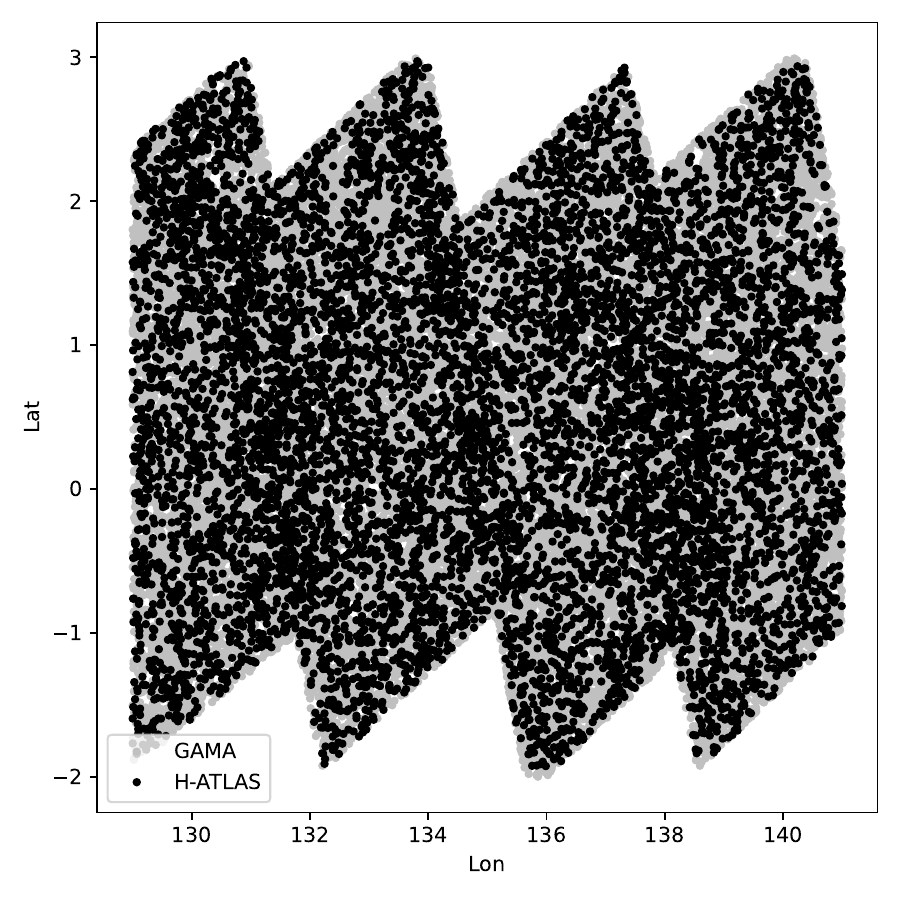}
    \includegraphics[width=0.15\textwidth]{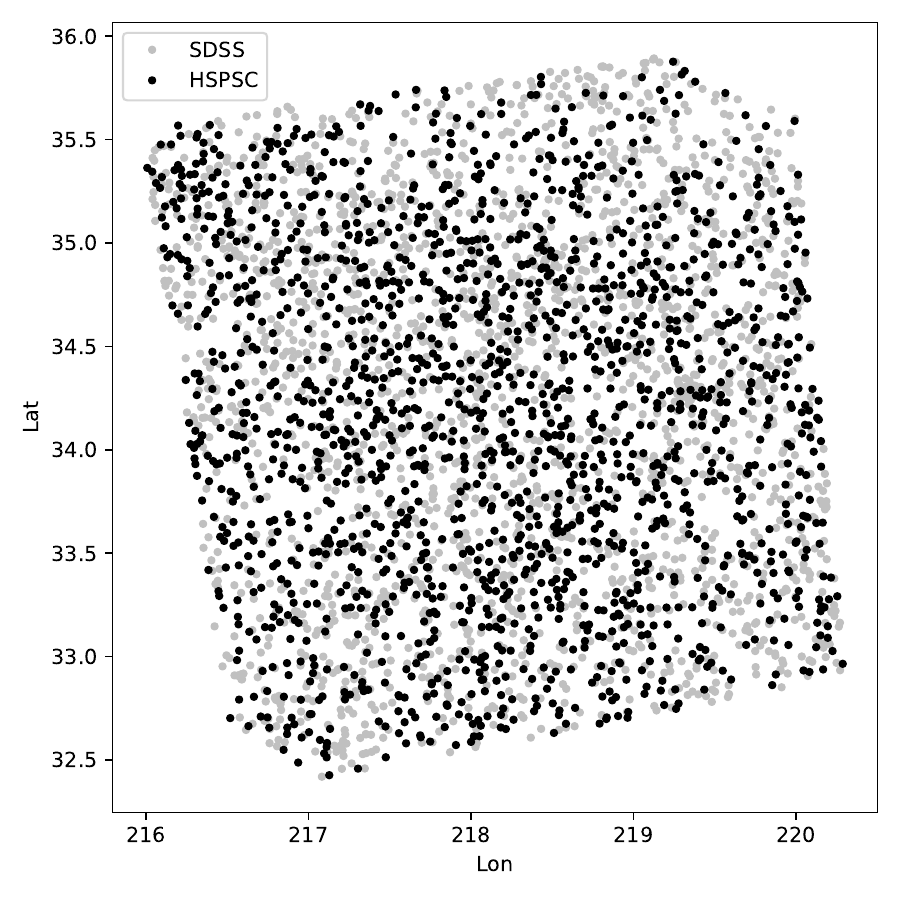}
    \includegraphics[width=0.15\textwidth]{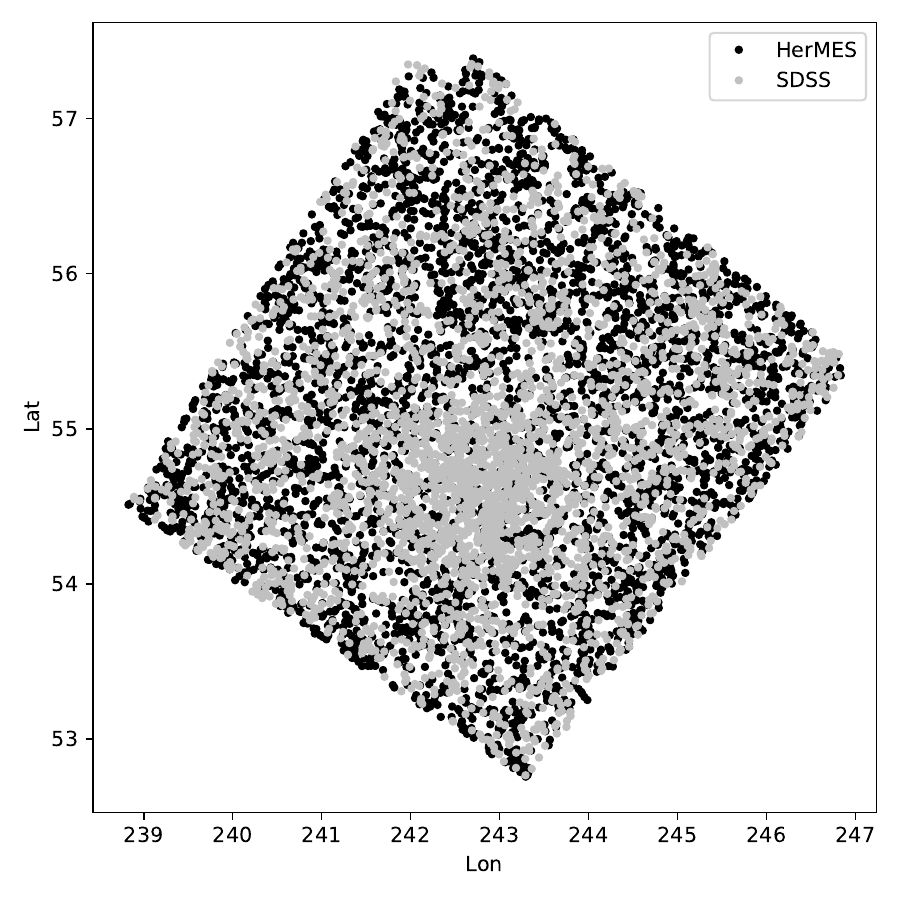}
    \includegraphics[width=0.15\textwidth]{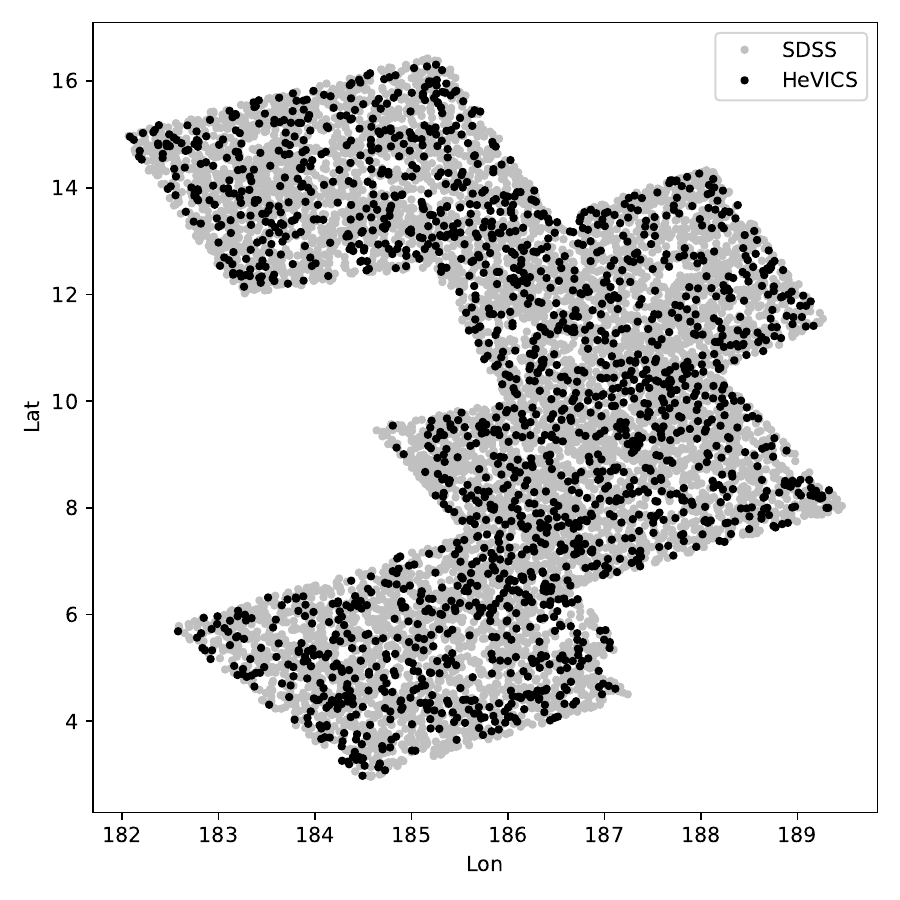}
    \includegraphics[width=0.15\textwidth]{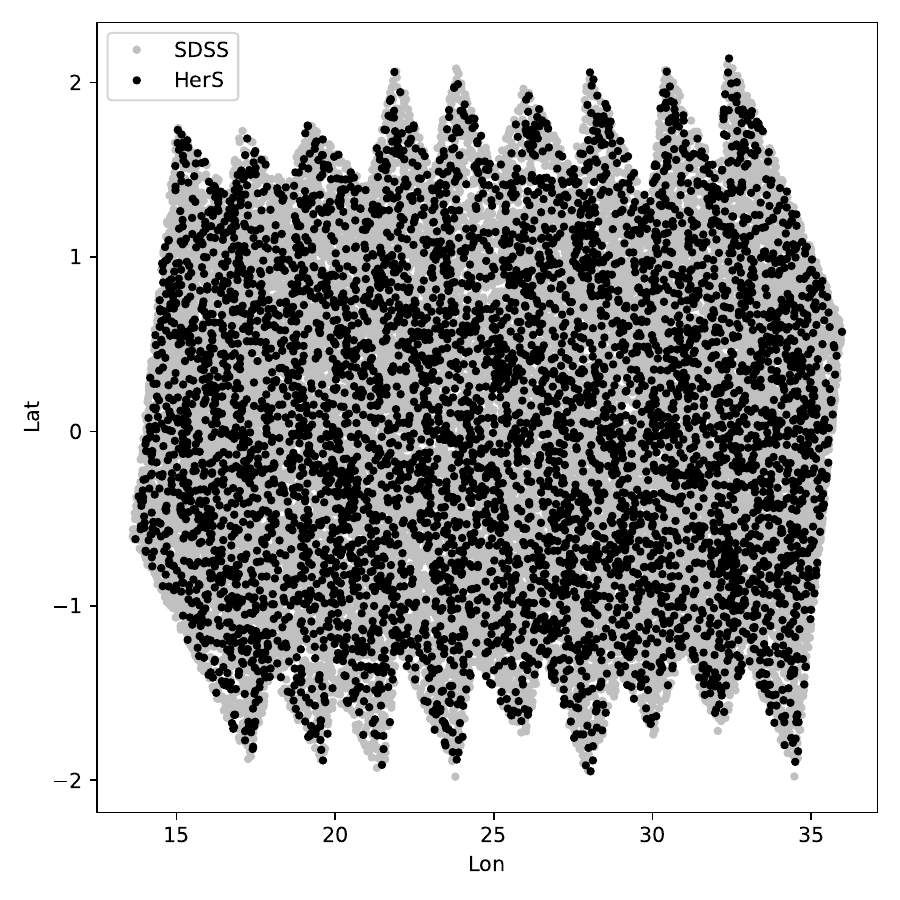}
    
    \caption{Examples of the final selection in each zone used for CCF. From left to right: H-ATLAS (GAMA-9), HSPSC (Boötes), HerMES (XMM), HeVICS, and HerS. Sources (SMGs) are shown in black, and lenses (GAMA or SDSS galaxies) in grey.
    }
    \label{fig:zone-examples}
\end{figure*}

Fig.~\ref{fig:zone-examples} illustrates the final selection for each catalogue. Lenses (GAMA or SDSS) are shown in grey over the source galaxies. The H-ATLAS sample (GAMA-09) is the largest and most homogeneous; scanning inhomogeneities are correctable via the noise map. The HSPSC zone (Boötes) shows density variations, notably an underdensity in the top right; further cleaning would reduce the area below viability. The HerMES XMM zone includes spurious central overdensity probably due to target mixing from different programs, and trimming it would leave too little area.

\clearpage
\section{Corner plots}
\label{app:corner_plots}

\begin{figure}[h]
  \centering
  \includegraphics[width=1\textwidth]{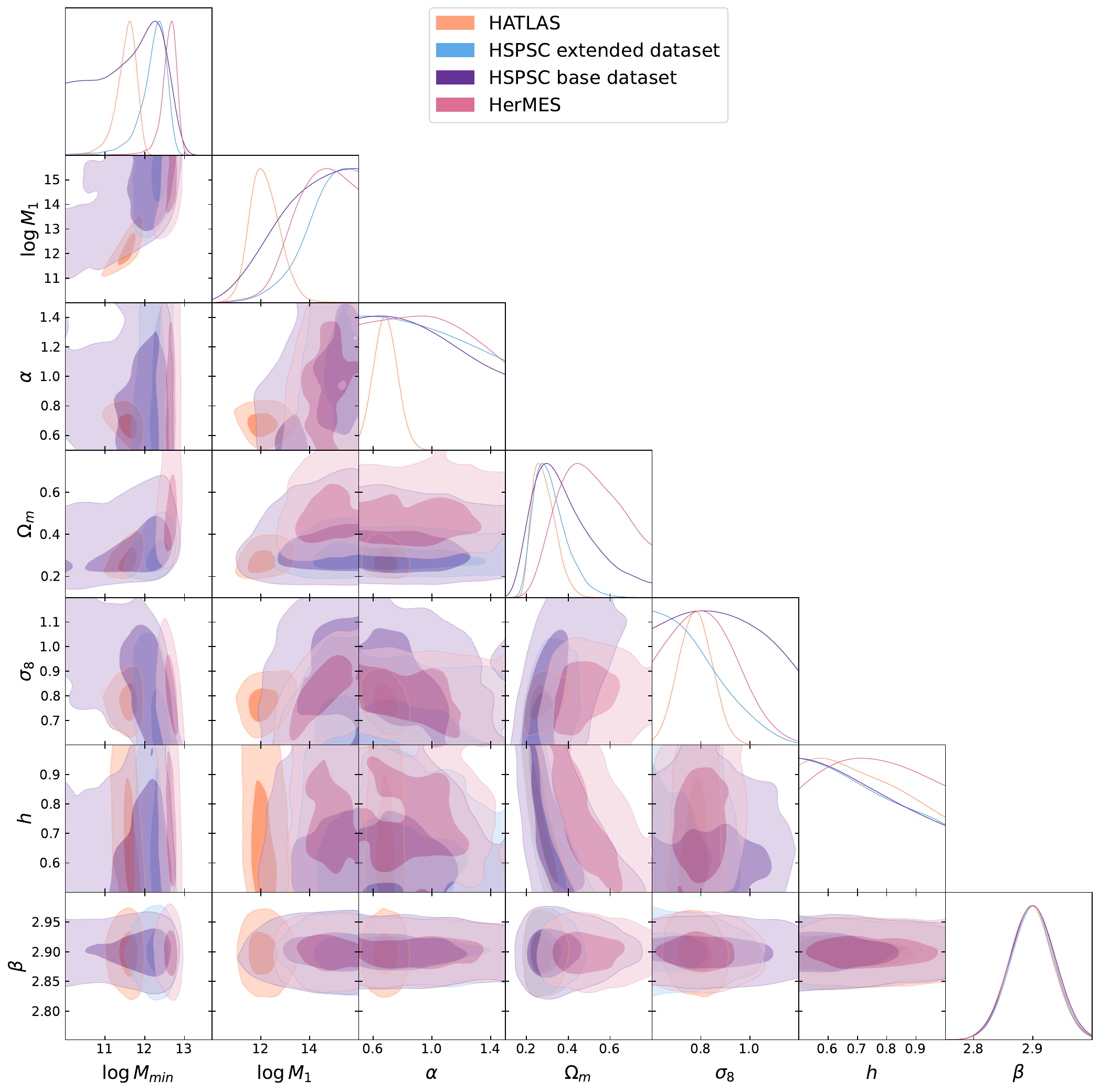}
  \caption{Marginalised posterior distributions and probability contours for the astrophysical and cosmological parameters using the HSPSC extended (light blue), HSPSC base (purple), and HerMES (pink) source datasets. In each case, overlapping SDSS regions have been employed as lenses. GAMA/H-ATLAS results are depicted in orange for comparison.
  }
  \label{fig:newcats-fullcorner}
\end{figure}

The full corner plots showing the marginalised posterior distributions and probability contours from the various analyses in this work are presented below. Fig. \ref{fig:HATLAS-fullcorner} shows the results of the proof-of-concept run, which used H-ATLAS as the background source sample and overlapping GAMA galaxies as lenses. Although most parameters show little dependence between astrophysical and cosmological quantities, an interesting degeneracy is observed between the HOD parameters and $\sigma_8$, particularly visible with $\alpha$ in the reference cases from \citet{CUE24}, and with $\log M_{\rm min}$ in H-ATLAS. Fig. \ref{fig:newcats-fullcorner} shows the results for the HSPSC (extended and base) and HerMES datasets. Due to the wider constraints, degeneracies bewtween astrophysical and cosmological parameters are not evident.   

\begin{figure}[h]
  \centering
  \includegraphics[width=1\textwidth]{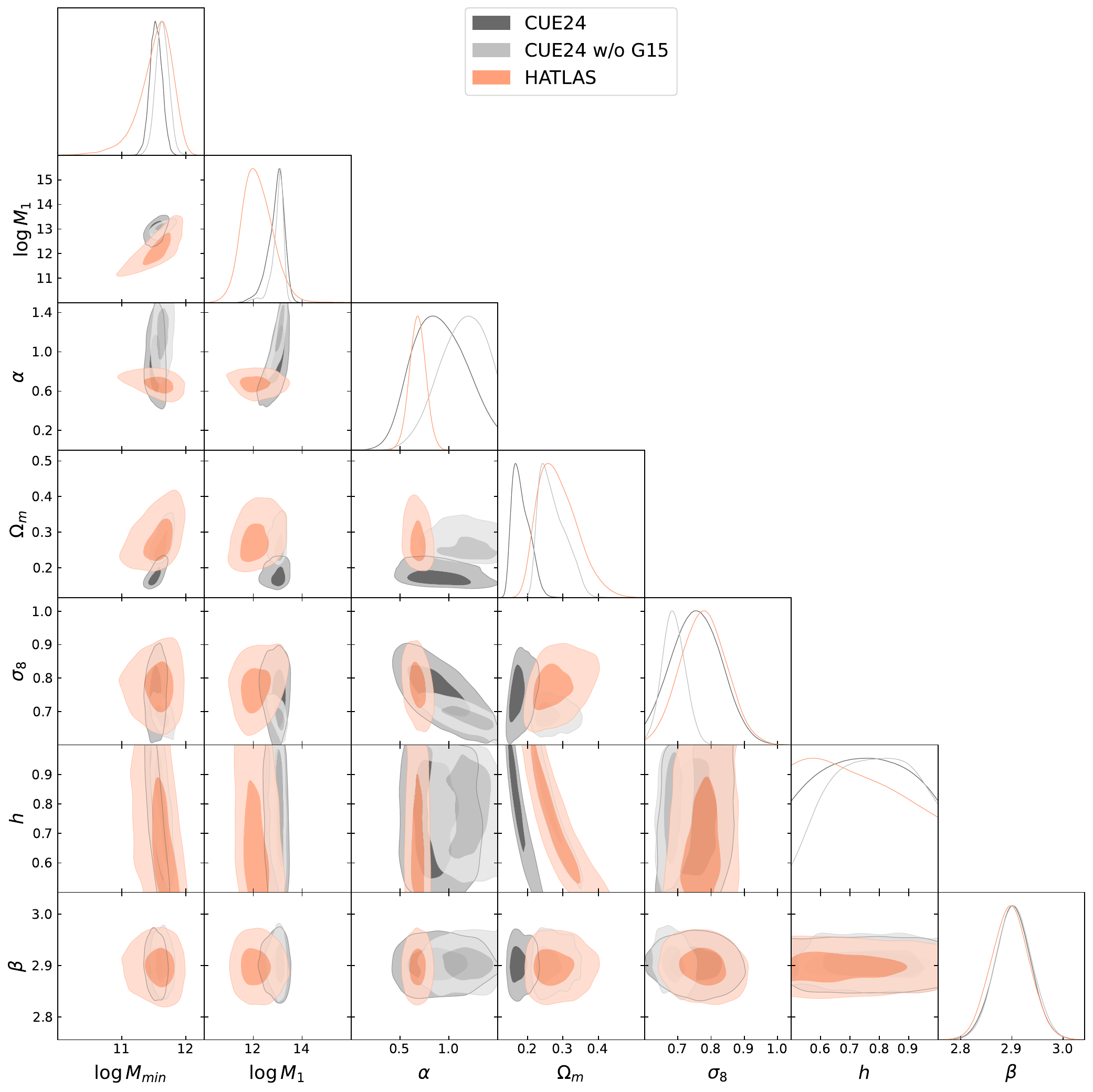}
  \caption{Marginalised posterior distributions and probability contours for the astrophysical and cosmological parameters in the GAMA/H-ATLAS datasets. The CCF derived using KDE for the random catalogue is displayed in orange, whereas the results from \citet{CUE24} are shown in dark grey (including all four sky regions: G09, G12, G15, and SGP) and in light grey (with G15 excluded from the analysis).
  }
  \label{fig:HATLAS-fullcorner}
\end{figure}

\clearpage

\section{Data tables}
\label{app:data_tables}
The full data constraints on the astrophysical and cosmological parameters from the various analyses in this work are presented below. 


\begin{table*}[h]
  \caption{
  Best-fit parameters and  68\% confidence interval determined using GAMA and H-ATLAS catalogues as lenses and sources, respectively. 
  }
  \label{tab:hatlas_results}
  \centering
  \begin{tabular}{cccccccccc} 
    \hline\hline
    \multicolumn{1}{c}{} & \multicolumn{3}{c}{HATLAS} & \multicolumn{3}{c}{CUE24} &
    \multicolumn{3}{c}{CUE24 w/o G15}\\
    \cmidrule(r){2-4} \cmidrule(r){5-7} \cmidrule(r){8-10}
    Parameter & Mean & Mode & 68\% CI & Mean & Mode & 68\% CI & Mean & Mode & 68\% CI \\
    \midrule
    $\log{M_\text{min}}$ & 
       11.52 & 11.62 & [11.14 , 11.83] &
       11.53 & 11.52 & [11.44 , 11.63] &
       11.63 & 11.62 & [11.52, 11.73]  \\
    $\log{M_{1}}$ &
       12.19 & 11.98 & [11.49 , 12.71] &
       12.93 & 13.07 & [12.71 , 13.33] &
       13.00 & 13.11 & [12.88, 13.27] \\
    $\alpha$ &
       0.68 & 0.68 & [0.60 , 0.76] &
       0.90 & 0.83 & [0.60 , 1.16] &
       1.12 & 1.20 & [0.96, 1.43] \\
    $\Omega_m$ & 
       0.29 & 0.26 & [0.22 , 0.33] &
       0.18 & 0.17 & [0.15 , 0.20] &
       0.27 & 0.24 & [0.22, 0.30] \\
    $\sigma_8$ &
       0.78 & 0.78 & [0.71 , 0.85] &
       0.76 & 0.76 & [0.68 , 0.83] &
       0.69 & 0.68 & [0.65, 0.72] \\
    \textit{h} &
       0.73 & - & [0.50 , 0.81] &
       0.75 & - & [0.50 , 1.00] &
       0.77 & 0.83 & [0.65, 0.94] \\
    $\beta$ & 
       2.90 & 2.90 & [2.86 , 2.93] &
       2.90 & 2.90 & [2.87 , 2.94] &
       2.90 & 2.90 & [2.87, 2.94] \\
    \hline\hline
  \end{tabular}
  \tablefoot{Left: random obtained using KDE. Middle and right: results from \citet{CUE24} employing the four sky zones G09, G12, G15 and SGP (CUE24) or excluding G15 from the analysis (CUE24 w/o G15). }
\end{table*}


\begin{table*}[h]
  \caption{Best-fit parameters and 68\% confidence interval using different source catalogues.}
  \label{tab:newcats_results}
  \centering
  \begin{tabular}{cccccccccc}
    \hline\hline
    \multicolumn{1}{c}{} & \multicolumn{3}{c}{HSPSC extended} & \multicolumn{3}{c}{HSPSC base} & \multicolumn{3}{c}{HerMES}\\
    \cmidrule(r){2-4} \cmidrule(r){5-7} \cmidrule(r){8-10} 
    Parameter  & Mean & Mode & 68\% CI & Mean & Mode & 68\% CI & Mean & Mode & 68\% CI \\
    \midrule
    $\log{M_{\text{min}}}$ 
    & 12.20 & 12.37 & [12.03 , 12.61] 
    & 11.56 & 12.25 & [10.94 , 12.66] 
    & 12.58 & 12.68 & [12.48 , 12.83] \\
    $\log{M_{1}}$ 
    & 14.63 & - & [14.31 , 16.00] 
    & 13.94 & - & [13.31 , 16.00] 
    & 14.36 & 14.66 & [13.81 , 15.90] \\
    $\alpha$
    & 0.97 & - & [0.50 , 1.50]
    & 0.95 & - & [0.50 , 1.10]
    & 0.98 & - & [0.50 , 1.50] \\
    $\Omega_m$
    & 0.32 & 0.28 & [0.22 , 0.37]
    & 0.39 & 0.30 & [0.20 , 0.47]
    & 0.51 & 0.44 & [0.34 , 0.64] \\
    $\sigma_8$
    & 0.78 & - & [0.60 , 0.82]
    & 0.88 & 0.81 & [0.63 , 1.01]
    & 0.82 & 0.80 & [0.65 , 0.92] \\
    $h$
    & 0.72 & - & [0.50 , 0.80]
    & 0.72 & - & [0.50 , 0.80]
    & 0.75 & - & [0.50 , 1.00] \\
    $\beta$
    & 2.90 & 2.90 & [2.86 , 2.94]
    & 2.90 & 2.90 & [2.86 , 2.94]
    & 2.90 & 2.90 & [2.86 , 2.94] \\
    \hline\hline
  \end{tabular}
\end{table*}


\begin{table*}[ht]
  \caption{Best-fit parameters and 68\% confidence interval using different source catalogues.}
  \label{tab:joint_results}
  \centering
  \begin{tabular}{cccccccccc}
    \hline\hline
    \multicolumn{1}{c}{} & \multicolumn{3}{c}{HATLAS} & \multicolumn{3}{c}{JOINTx2} & \multicolumn{3}{c}{JOINTX4}\\
    \cmidrule(r){2-4} \cmidrule(r){5-7} \cmidrule(r){8-10} 
    Parameter  & Mean & Mode & 68\% CI & Mean & Mode & 68\% CI & Mean & Mode & 68\% CI \\
    \midrule
    
    $\Omega_m$ & 
       0.29 & 0.26 & [0.22 , 0.33] &
       0.27 & 0.24 & [0.21 , 0.31] &
       0.30 & 0.27 & [0.23, 0.35] \\
    $\sigma_8$ &
       0.78 & 0.78 & [0.71 , 0.85] &
       0.72 & 0.70 & [0.63 , 0.78] &
       0.80 & 0.79 & [0.73, 0.87] \\
    $h$ &
       0.73 & - & [0.50 , 0.81] &
       0.72 & - & [0.50 , 0.80] &
       0.72 & - & [0.50 , 0.80] \\
    
    \hline\hline
  \end{tabular}
\end{table*}

\end{appendix}
\end{document}